\RequirePackage{pdf14}

\documentclass[letterpaper]{article}
\usepackage[usenames,dvipsnames,table]{xcolor}
\usepackage{graphicx}
\usepackage{caption}
\usepackage{subcaption}
\usepackage{array,setspace,mathrsfs,amsthm,amsfonts,amsmath,yfonts,dsfont,bbm,colonequals,mathtools}
\usepackage{tikz-cd}

\usepackage{jheppub}
\usepackage{afterpage}
\usepackage{xspace}
\usepackage{stmaryrd}
\usepackage[normalem]{ulem}
\usepackage{tikz}
\usetikzlibrary{decorations.markings}

\newcommand{\beq}{\begin{equation}}
\newcommand{\eeq}{\end{equation}}

\newcommand\nn{\nonumber}
\newcommand\ph{\phantom}

\newcommand{\cA}{\mathcal{A}}
\newcommand{\cB}{\mathcal{B}}
\newcommand{\cC}{\mathcal{C}}
\newcommand{\cD}{\mathcal{D}}

\newcommand{\cF}{\mathcal{F}}
\newcommand{\cI}{\mathcal{I}}
\newcommand{\cJ}{\mathcal{J}}
\newcommand{\cL}{\mathcal{L}}
\newcommand{\cM}{\mathcal{M}}
\newcommand{\cN}{\mathcal{N}}
\newcommand{\cO}{\mathcal{O}}
\newcommand{\cQ}{\mathcal{Q}}
\newcommand{\cR}{\mathcal{R}}
\newcommand{\cS}{\mathcal{S}}
\newcommand{\cW}{\mathcal{W}}

\newcommand\qq{\mathbbmtt{Q}}

 \newcommand{\g}{\mathfrak{g}}
  \newcommand{\su}{\mathfrak{su}}

\title{Supersymmetric Localization in AdS$_5$ and the Protected Chiral Algebra
 }
\author[1]{Federico Bonetti,\!}
\author[2]{Leonardo Rastelli\ph{,}\!}

\emailAdd{federico.bonetti@stonybrook.edu}
\emailAdd{leonardo.rastelli@stonybrook.edu}

\affiliation[1,2]{C.~N.~Yang Institute for Theoretical Physics, Stony Brook University, Stony Brook, NY 11794, USA}

\bigskip
\abstract{
${\cal N} =4$ super Yang-Mills theory 
 admits \cite{Beem:2013sza} a protected subsector isomorphic to a two-dimensional  chiral algebra, obtained by passing to the cohomology of a certain  supercharge.
 In the large $N$ limit, we expect this chiral algebra to have a dual description as a subsector of   IIB supergravity on $AdS_5 \times S^5$. This subsector can be carved out by  a version of supersymmetric localization, using the bulk analog of the boundary  supercharge. We illustrate this procedure in a simple model, the theory of an ${\cal N}=4$ vector multiplet in  $AdS_5$, for which a convenient off-shell description is available. This model can be viewed  as the simplest truncation of  the full $AdS_5 \times S^5$ supergravity,
 in which case the vector multiplet should be taken in the adjoint representation of ${\mathfrak g}_F = \mathfrak {su}(2)_F$.   Localization
  yields Chern-Simons theory on $AdS_3$ with gauge algebra ${\mathfrak g}_F$, whose boundary dual is the affine Lie algebra $\widehat {\mathfrak g}_F$.
  We comment on the generalization to the full bulk theory. We propose that the large $N$ limit of the chiral algebra of ${\cal N}=4$ SYM is again dual to Chern-Simons theory, with   gauge algebra a suitable higher-spin superalgebra. 
}


\begin{document}
\setcounter{tocdepth}{2}
\maketitle
\setcounter{page}{1}


\section{Introduction}
\label{sec:intro}

Any four-dimensional ${\cal N}=2$ superconformal field theory (SCFT) admits a subsector of  correlation functions that exhibits the structure of a two-dimensional
chiral algebra \cite{Beem:2013sza}.   This is in particular the case for ${\cal N}=4$ super Yang-Mills (SYM) theories. The associated chiral algebra is
labelled  by the  gauge algebra $\mathfrak{g}$  and is independent of the complexified gauge coupling. It
encodes an infinite amount of  information about a very rich protected subsector of the SYM theory.
In this paper we start addressing the  question
of finding a holographic description  of  this protected chiral algebra for $\g = \su(N)$,  in the  large $N$ limit. Answering this question would provide us with a new solvable model  of holography.
Rather than a mere toy example, this would be an intricate yet tractable model carved out naturally
  from the standard holographic duality.

While in the general  ${\cal N}=2$ case the protected chiral algebra has no residual supersymmetry, the chiral algebra  associated to an ${\cal N}=4$ SCFT  contains
the {\it small} ${\cal N}=4$ superconformal algebra (SCA) as a subalgebra.  
Conjecturally  \cite{Beem:2013sza}, the chiral algebra for ${\cal N}=4$ SYM theory with gauge algebra $\g$ is 
a novel ${\cal N}=4$ super chiral algebra, strongly generated by a finite number of currents. The  super chiral algebra generators descend from the generators of the one-half BPS chiral ring of the SYM theory, and are thus in one-to-one correspondence with the Casimir invariants of $\g$.
For example, for $\g = \su(N)$,  the super chiral algebra is conjectured to have $N-1$  generators, of  holomorphic dimension $h = 1,  \frac{3}{2}, \dots \frac{N}{2}$, in correspondence with
the  familiar single-trace one-half BPS operators of the SYM theory,  namely ${\rm Tr} \, X^{2h}$ 
 in the symmetric traceless representation of the $\mathfrak{so}(6)$ R-symmetry. As we will review in detail below, only an $\su(2)_F$ subalgebra of $\mathfrak{so}(6)$ is visible in the chiral algebra,
 where it is in fact enhanced to  the affine Kac-Moody algebra  $\widehat {\mathfrak{su}(2)}_F$ that is part of the
small  ${\cal N}=4$ SCA. The super chiral algebra generator of dimension $h$ transforms in the spin $h$ representation of $\mathfrak{su}(2)_F$.
This is a BPS condition -- the generators with $h >1$ are the highest-weight states of {\it short} representations 
 of the ${\cal N}=4$ subalgebra.\footnote{Other examples of  ${\cal N}=4$ super chiral algebra
 have been considered in the literature, see {\it e.g.}, \cite{Beccaria:2014jra}, where the generators are taken to be $\su(2)_F$  singlets  and are thus highest-weight states of {\it long} representations of the ${\cal N}=4$ SCA.}
The central charge of the chiral algebra is given by $c_{\rm 2d} = - 3\,{\rm dim}\, \mathfrak{g}=
-3 (N^2 -1)$. It is not known whether the chiral algebra for fixed $N > 2$ admits a deformation\footnote{For $N=2$, the chiral algebra coincides with the small ${\cal N}=4$ SCA, which is  of course consistent for any central charge.} to general values of the central charge. In fact,  for this special value of $c_{\rm 2d}$ one finds several null relations that  might be essential
to ensure associativity of the operator algebra.

Let us now consider the large $N$ holographic description.    As familiar,  ${\cal N}=4$ SYM theory is dual to IIB string theory on $AdS_5 \times S^5$, with the $1/N$ expansion on the field theory side
corresponding to the topological expansion on the string theory side.
It would be extremely interesting to construct  a  ``topological"  string theory whose genus expansion reproduces
the $1/N$ expansion of the ${\cal N}=4$ SYM chiral algebra. 
Here we will address the simpler question of finding a holographic description for the leading large $N$ limit of the chiral algebra,
in terms of a classical field theory in the bulk.  There are two ways we can imagine to proceed: attempting to construct the bulk theory by bottom-up guesswork; or deriving it  from the top-down as a subsector 
of $AdS_5 \times S^5$ string field theory.

 From the bottom-up perspective, the natural conjecture is that the bulk theory is a Chern-Simons field theory in $AdS_3$, with gauge algebra a suitable
infinite-dimensional supersymmetric higher-spin algebra. Such a duality would mimic several examples of ``higher-spin holography" that have been studied in recent years
in the context of the $AdS_3/{\rm CFT}_2$ correspondence. A duality
has been proposed in \cite{Gaberdiel:2010pz}\footnote{This proposal, reviewed  in \cite{Gaberdiel:2012uj},
 has  
 passed several tests \cite{Gaberdiel:2011zw,Gaberdiel:2011wb}
and has been extended in various directions,
both in a purely bosonic context \cite{Gaberdiel:2011nt, Candu:2012ne}
and in models with supersymmetry, see for instance
 \cite{Candu:2012jq, Creutzig:2012ar,Gaberdiel:2013vva,Beccaria:2013wqa, Creutzig:2014ula,
 Candu:2014yva}.}
 between 
 higher-spin Vasiliev theory in $AdS_3$ \cite{Prokushkin:1998bq} and 
a suitable   't Hooft limit  
of  $\cW_N$ minimal models, {\it i.e.}~the coset CFTs $\mathfrak {su}(N)_k \otimes \mathfrak {su}(N)_1 /  \mathfrak {su}(N)_{k+1}$. 
  In this example, the chiral algebra that controls
the large $N$ limit is  the ${\cal W}_\infty[\mu]$ algebra, where the parameter $\mu$ is identified with the 't Hooft coupling $N/(N+k)$, kept fixed as $N \to \infty$. The bulk dual description involves  Chern-Simons theory with gauge algebra
the infinite-dimensional Lie  algebra $\mathfrak{hs}[\mu]$. The (non-linear) ${\cal W}_\infty[\mu]$ algebra arises at the asymptotic symmetry of this Chern-Simons theory \cite{Henneaux:2010xg,Campoleoni:2010zq,Campoleoni:2011hg}.
We find it likely that our example will work along similar lines, 
but we have not yet been able to identify the correct supersymmetric higher-spin algebra. An obvious feature of the sought after higher-spin algebra is that it must
 contain $\mathfrak{psu}(1, 1|2)$ as a subalgebra. Indeed the asymptotic symmetry of $AdS_3$ Chern-Simons theory with algebra  $\mathfrak{psu}(1, 1|2)$ is the small ${\cal N}=4$ SCA (see for instance \cite{Ito:1992bi,deBoer:1998kjm} and references therein), which, as reviewed above, is a consistent truncation of the full super W-algebra. 
  In our case, 
the construction of the complete higher-spin algebra is made  more challenging by the absence of an obvious deformation parameter analogous to the 't Hooft coupling of the ${\cal W}_N$ minimal models\footnote{Recall that the usual 't Hooft coupling $g_{YM}^2 N$  is not visible in the chiral algebra, which describes a protected subsector of observables of the SYM theory.} -- as we have remarked, the chiral algebra for $\su(N)$ SYM theory might be  isolated, stuck at  a specific value of the central charge.

The top-down approach  is conceptually straightforward.
The dual bulk theory must be a subsector of   IIB  {\it supergravity} on $AdS_5 \times S^5$. Indeed,  the  generators of the chiral algebra descend from the single-trace one-half BPS operators
of ${\cal N}=4$ SYM, which are dual to 
 the infinite tower of Kaluza-Klein (KK) supergravity modes on  $S^5$.     In principle, our task is clear.
 In the boundary SYM theory, the $2d$ chiral subsector is carved out by passing to the cohomology of either one of two nilpotent supercharges  \cite{Beem:2013sza}. 
 The bulk supergravity admits analogous nilpotent supercharges.  We then expect to find the bulk dual to the large $N$ limit of the chiral algebra by 
  localization of the supergravity theory  with respect to either supercharge. 
   In practice however, this program is difficult to implement rigorously. The  technique of supersymmetric localization
requires an off-shell formalism, but we are not aware of  such a formalism for $AdS_5 \times S^5$ supergravity, or even for its consistent truncation to  ${\cal N}=8$ $AdS_5$ supergravity.

In this paper, we give a proof of concept that this localization program works as expected, producing  an $AdS_3$ Chern-Simons theory out of  $AdS_5$ supergravity.
We consider the simplest truncation of the supergravity theory for which a convenient off-shell formalism is readily available: the theory of an ${\cal N}=4$ vector multiplet in $AdS_5$,
 covariant under an $\mathfrak{su}(2, 2 |2)$ subalgebra of the full $\mathfrak{psu}(2, 2 |4)$ superalgebra.
We obtain this model by a straightforward analytic continuation of the analogous model on $S^5$ \cite{Minahan:2015jta}. When viewed as part of the  ${\cal N}=8$ supergravity multiplet, the ${\cal N}=4$ vector multiplet
transforms in the adjoint representation of $\su(2)_F$ (the centralizer of the embedding $\mathfrak{su}(2, 2 |2) \subset \mathfrak{psu}(2, 2 |4)$),  
 but it is no more difficult to  consider 
 a general simple Lie algebra $\mathfrak{g}_F$. We show by explicit calculation that supersymmetric localization with respect to the relevant
 supercharge yields Chern-Simons theory in $AdS_3$, with gauge algebra ${\mathfrak{g}}_F$, and level $k$ related to the Yang-Mills coupling.  As is well-known, its dual boundary theory is  the affine Kac-Moody algebra  
 $\widehat {\mathfrak{g}}_F$ at level $k$.  Apart from confirming the general picture that we have outlined, we believe that the details of our calculations are interesting in their own right, and may find a broader range
 of applications. 
Localization computations involving non-compact AdS backgrounds
have been considered in the literature, see for instance \cite{Dabholkar:2010uh,Dabholkar:2011ec,Gupta:2012cy,Murthy:2013xpa,Dabholkar:2014wpa,Murthy:2015yfa,Gupta:2015gga,Murthy:2015zzy,Aharony:2015hix}
and more recently \cite{Assel:2016pgi}. It is worth pointing out that the 
Killing spinor used in 
our localization computation satisfies somewhat unusual algebraic
properties compared
to those usually assumed in past work. This is a consequence of the fact that
our choice of supercharge mimics the (somewhat unusual) cohomological construction
on the field theory side.

 Localization of the full maximally supersymmetric $AdS_5$ supergravity would be technically challenging, but it seems very plausible (by supersymmetrizing the above result) that it would yield $AdS_3$ Chern-Simons theory with gauge algebra $\mathfrak{psu}({1, 1 |2})$, whose boundary dual is the small ${\cal N}=4$ superconformal algebra. Inclusion of the KK modes is however much harder, and at present the quest
 for the full holographic dual seems best pursued  by bottom-up guesswork of the higher-spin superalgebra.

 The rest of the paper is organized as follows. In section 2 we review the construction and main features of the chiral algebra associated to an ${\cal N}=2$ SCFT. Section 3 contains our main result,
 the localization of the ${\cal N}=4$ super Yang-Mills action in $AdS_5$ to bosonic Chern-Simons theory in $AdS_3$. In section 4 we collect some useful facts and offer some speculations for the construction of
  the full holographic dual  of the ${\cal N}=4$ SYM chiral algebra. We conclude in section 5 with a brief discussion. An appendix contains conventions and technical material.


\section{Review of the chiral algebra construction}
\label{sec:review}

In an effort to make this paper self-contained, we briefly  review in this section the construction of the two-dimensional
chiral algebra associated to a four-dimensional $\cN = 2$ superconformal field theory \cite{Beem:2013sza}. 
Our main focus
is on ${\cal N}=4$ SYM theory, but the calculations  of section 3 will be relevant for  any ${\cal N}=2$ SCFT that admits a supergravity dual and enjoys
a global symmetry. To this end, we review in   section 2.2 the special properties of chiral algebras associated to  SCFTs with additional (super)symmetries.

\subsection{Cohomological construction}
The spacetime signature and the reality properties of 
operators are largely inessential in the following. We thus consider
the complexified theory on flat $\mathbb C^4$. 
The complexified superconformal algebra is $\mathfrak{sl}(4|2)$,
with maximal bosonic subalgebra 
$\mathfrak {sl}(4) \oplus \mathfrak {sl}(2)_R \oplus \mathbb C_r$.
The first term corresponds to the action of the complexified 
conformal algebra on $\mathbb C^4$,
while the other terms constitute the complexification of the R-symmetry
of the theory.

Let us consider a fixed (complexified) plane $\mathbb C^2 \subset \mathbb C^4$.
The subalgebra of the conformal algebra preserving 
the plane is of the form
\beq
\mathfrak {sl}(2) \oplus \overline {\mathfrak {sl}(2)} \oplus \mathbb C_\perp
\oplus \subset \mathfrak {sl}(4) \ .
\eeq
The first two summands comprise the complexified
two-dimensional conformal algebra acting on the fixed plane,
while $\mathbb C_\perp$ corresponds to complexified rotations
in the two directions orthogonal to the plane.
We use the notation $L_n$, $n=-1,0,1$ for the generators of the first
summand, and $\overline L_n$ for the second summand,
while the generator of $\mathbb C_\perp$ will be denoted
$\cM_\perp$.
It is natural to adopt coordinates $\zeta$, $\bar\zeta$
on the selected plane $\mathbb C^2$,
with $\mathfrak {sl}(2)$ acting on $\zeta$
via M\"obius transformations, and   
 $ \overline {\mathfrak {sl}(2)}$ acting similarly on   $\bar \zeta$.
 
The R-symmetry of the superconformal theory
allows us to define a suitable diagonal subalgebra
\beq
\widehat{\mathfrak{sl}(2)} \subset  \overline {\mathfrak {sl}(2)} 
\oplus \mathfrak {sl}(2)_R \ ,
\eeq
given explicitly by
\beq
\widehat L_{-1} = \overline L_{-1} + \cR^- \ , \qquad
\widehat L_{0} = \overline L_{} - \cR  \ , \qquad
\widehat L_{1} = \overline L_{1} - \cR^+  \ ,
\eeq
where $\cR$, $\cR^\pm$ denote   the generators of $\mathfrak {sl}(2)_R$
with commutators
\beq
[\cR^+ , \cR^-] = 2\, \cR \ , \qquad
[\cR , \cR^\pm] = \pm \cR^\pm \ .
\eeq
The relevance of the twisted subalgebra $\widehat{\mathfrak{sl}(2)}$
stems from the following crucial fact. 
There exist 
two linear combinations $\qq_{\,1}$, $\qq_{\,2}$
of the supercharges of $\mathfrak {sl}(4|2)$,
inequivalent under similarity transformations,
that enjoy the following properties:
\beq
\begin{array}{l}
\begin{array}{ccccccc}
\{ \qq_{\, 1} , \qq_{\, 1} \} & =&  0  \ , &\phantom{m}& \{ \qq_{\, 2} , \qq_{\, 2} \} & =&  0 \ ,  \\[2mm]
{} [ L_n,  \qq_{\, 1} ] & = &  0 \ ,&   &  [ L_n,  \qq_{\, 2} ] & = &  0 \ ,
\end{array}
\\[6mm]
\; \widehat L_n = \{ \qq_{\,1} , \cF_{1,n} \} =  \{ \qq_{\,2} , \cF_{2,n} \}  \ , \qquad \;\;\;
\text{for suitable odd generators $\cF_{1,n}$, $\cF_{2,n}$ of $\mathfrak {sl}(4|2)$ } \ ,
\\[2mm]
\; \{ \qq_{\, 1} , \qq_{\, 2} \} = - r - \cM_\perp \ ,
\end{array}
\eeq
where $r$ denotes the generator of $\mathbb C_r$.
In other words, the supercharges  $\qq_{\,1}$, $\qq_{\,2}$
are nilpotent, are invariant under the action of the  holomorphic
factor $\mathfrak {sl}(2)$ of the conformal algebra of the plane,
and are such that the twisted antiholomorphic factor 
$\widehat {\mathfrak {sl}(2)}$ is both a $\qq_{\,1}$- and a $\qq_{\,2}$-commutator.
Explicit expressions for $\qq_{\, 1}$, $\qq_{\,2}$ in a convenient
basis are found in \cite{Beem:2013sza}, where it is also
shown that
\beq
[ \widehat L_0 , \qq_{\, i} ] = 0 \ , \qquad 
[r + \cM_\perp , \qq_{\, i} ] = 0 \  , \qquad i = 1,2 \ .
\eeq 
The chiral algebra associated to the four-dimensional superconformal
field theory is then defined by considering cohomology classes of 
operators with respect to $\qq_{\, i}$ ($i = 1$ or 2), {\it {\it i.e.}}~the 
set of operators (anti)commuting with $\qq_{\, i}$
modded out by addition of arbitrary $\qq_{\, i}$-commutators.

Let $\cO$ be a local operator of the four-dimensional theory
such that its insertion at the origin 
 $\cO(0)$ 
defines a non-trivial $\qq_{\,i}$ cohomology class
($i = 1$ or $2$), {\it {\it i.e.}}~$[ \qq_{\, i} , \cO(0) \} = 0$, 
but $\cO(0)$ is not itself a $\qq_{\, i}$-commutator. It follows that
$\cO(0)$ necessarily commutes with 
  $ \widehat L_0$ and $r + \cM_\perp$. In terms of the four dimensional quantum
numbers of $\cO$, this amounts to
\beq \label{Schur_restrictions}
\tfrac 12 ( \Delta - j_1 - j_2)  - R  = 0 \ , \qquad
r + j_1 - j_2 = 0 \ ,
\eeq
where $\Delta$ is the conformal dimension of $\cO$, $j_1$, $j_2$ are its
Lorentz Cartan quantum numbers, $R$ is the $\mathfrak {sl}(2)_R$ Cartan
quantum number, and $r$ is the $\mathbb C_r$ quantum number.
The restrictions \eqref{Schur_restrictions} on the quantum numbers 
of a four dimensional operators define the so-called 
class of {\it Schur operators} of the theory.
A Schur operator is always the highest-weight
state of its Lorentz and R-symmetry multiplet.
If the latter is denoted schematically as
$\cO^{I_1 \dots I_{2R}} _{\alpha_1 \dots \alpha_{2j_1}
\dot \beta_1 \dots \dot \beta_{2j_2}}$,
where $I = 1,2$ are $\mathfrak{sl}(2)_R$ fundamental indices,
and $\alpha = +,-$, $\dot \beta = \dot + , \dot -$ are spacetime Weyl indices,
then the Schur operator is 
$\cO^{1 \dots 1}_{+ \dots+ \dot + \dots \dot +}$.
Let us also point out that,
if the theory is unitary, \eqref{Schur_restrictions} are not only
necessary but also sufficient conditions
for $\cO(0)$ to define a non-trivial $\qq_{\, i}$ cohomology class.
Furthermore, in that case  $\qq_{\, 1}$ and $\qq_{\, 2}$ define the same cohomology.
We refer the reader to  \cite{Beem:2013sza} for an explanation
of these points.

Suppose $\cO(0)$ defines a non-trivial $\qq_{\, i}$-cohomology class.
We cannot translate this operator away from the origin along the
directions orthogonal to the $(  \zeta, \bar \zeta)$ plane
without losing $\qq_{\, i}$-closure, since $\qq_{\, i}$ is not invariant
under translations in those directions.
We can, however, construct the following twisted
translated operator
\beq
\tilde \cO(\zeta , \bar \zeta) = e^{\zeta L_{-1} + \bar \zeta \widehat L_{-1}} \cO(0)
e^{-\zeta L_{-1} - \bar \zeta \widehat L_{-1}}  \ ,
\eeq
which is still annihilated by $\qq_{\,i}$. 
This object can also be written
as a $\bar \zeta$-dependent linear combination
of the R-symmetry components of the 
multiplet to which the Schur operator belongs,
\beq
\tilde \cO(\zeta , \bar \zeta)  = u_{I_1} (\bar \zeta) \dots
u_{I_{2R}} (\bar \zeta) \,
\cO^{I_1 \dots I_{2R}} _{+ \dots +
\dot + \dots \dot + } (\zeta , \bar \zeta) \ , \qquad
u_1(\bar \zeta) =1 \ , \qquad u_2(\bar \zeta) = \bar \zeta \ .
\eeq
Crucially, 
thanks to the fact that the generators $\widehat{ \mathfrak{sl}(2) }$
are  $\qq_{\,i}$-exact, the antiholomorphic dependence of 
$\tilde \cO(\zeta , \bar \zeta)$ is trivial in cohomology,
\beq
\partial_{\bar \zeta} \, \tilde \cO(\zeta , \bar \zeta)  = [\qq_{\,i} , \dots \}  \ . 
\eeq
This suggests the notation
\beq
\chi[\cO](\zeta) = \text{$\qq_{\,i}$-cohomology class of $\tilde \cO(\zeta , \bar \zeta)$} \ .
\eeq
Correlators  of cohomology classes $\chi[\cO](\zeta)$
are defined in terms of correlators
of the representatives  $\tilde \cO(\zeta , \bar \zeta)$,
\begin{align}
\langle \chi[\cO_1](\zeta_1) \dots \chi[\cO_n](\zeta_n) \rangle
&= \langle   \tilde \cO_1(\zeta_1 , \bar \zeta_1)
\dots  \tilde \cO_n(\zeta_n , \bar \zeta_n) \rangle  \ , 
\end{align}
are independent of the representative chosen, and depend meromorphically
on the insertion points.
By the same token, the four-dimensional OPE of
two 
 twisted
translated Schur operators  
$\tilde \cO_1$, $\tilde \cO_2$  induces a meromorphic OPE
of cohomology classes $\chi[\cO_1]$, $\chi[\cO_2]$.
Let us remark that the holomorphic dimension $h$ 
of $\chi[\cO]$ is given in terms of the quantum numbers of the
four-dimensional Schur operator $\cO$ by
\beq
h = \tfrac 12 (\Delta + j_1 + j_2) \ .
\eeq
This quantity is generically a half-integer,
but
as a consequence of four-dimensional $\mathfrak{sl}(2)_R$
selection rules, the OPE of any 
  two cohomology classes $\chi[\cO_1]$, $\chi[\cO_2]$
  is single-valued in the $\zeta$-plane.

\subsection{Affine enhancement of symmetries} \label{sec:affine_enhancement}

The stress tensor of a four-dimensional $\cN = 2$ theory
sits in a supersymmetry multiplet of type $\hat \cC_{0(0,0)}$
in the notation of \cite{Dolan:2002zh}. The same multiplet contains
the $\mathfrak{sl}(2)_R$ symmetry current of the theory,
$J^{(IJ)}_{\alpha \dot \beta}$. Its Lorents and R-symmetry
highest-weight component is a Schur operator and determines
an element of the chiral algebra with holomorphic dimension
two,
\beq
{\mathsf T} =  \chi [J^{11}_{+ \dot +}] \ .
\eeq
This object is identified with the stress tensor of the chiral algebra.\footnote{We
refer the reader to \cite{Beem:2013sza} for a careful
discussion of the relative normalization of
$\mathsf T$, $J^{11}_{+\dot +}$
in a standard set of conventions for 
four-dimensional and two-dimensional operators,
and similarly for other pairs of four-dimensional and
two-dimensional operators discussed below.
}
The meromorphic $\mathsf{TT}$ OPE is determined by the
OPE of R-symmetry currents in four dimensions,
and has the expected form with a two-dimensional
central charge
\beq \label{c2d_vs_c4d}
c_{\rm 2d} = -12 \, c_{\rm 4d} \ ,
\eeq
where $c_{\rm 4d}$ is one of the two conformal anomaly
coefficients of the four-dimensional theory \cite{Beem:2013sza}. 
Unitarity in four dimensions requires $c_{\rm 4d}>0$,
yielding a non-unitary chiral algebra in two dimensions.

If the four-dimensional theory is invariant under  a   continuous flavor symmetry group
$G_F$,
its spectrum contains a
conserved current in the adjoint of the flavor symmetry algebra $\mathfrak{g}_F$.
The latter is contained in a 
 supersymmetry multiplet of type $\hat \cB_1$,
which also includes an $\mathfrak{sl}(2)_R$ triplet of
scalars $M^{(IJ)}$ in the adjoint of $\mathfrak{g}_F$
with $\Delta = 2$.
The R-symmetry highest weight component of $M^{IJ}$
is a Schur operator, yielding an element of the chiral algebra with
holomorphic dimension one,
\beq
\mathsf J =   \chi [M^{11}] \ .
\eeq
The $\mathsf{JJ}$ OPE reveals that this object can be identified
with an affine current in two dimensions, satisfying a Kac-Moody
algebra based on the Lie algebra $\mathfrak g_F$
with level
\beq
k_{\rm 2d} = - \tfrac 12 k_{\rm 4d}  \ ,
\eeq
where $k_{\rm 4d}$ is an anomaly coefficient entering 
 the four-dimensional
OPE of two flavor currents \cite{Beem:2013sza}.

The cohomological construction of the previous section
can also be performed in theories with $\cN = 3$ or $\cN = 4$
superconformal symmetry. The spectrum of such a theory,
expressed in $\cN= 2$ language,
contains additional conserved spin $3/2$ supersymmetry currents.
The latter are contained in supermultiplets of type 
$\cD_{\scriptscriptstyle \frac 12(0,0)}$,
$\bar \cD_{\scriptscriptstyle \frac 12(0,0)}$,
which also include an $\mathfrak{sl}(2)_R$
triplet of spin $1/2$ operators $\Psi^{(IJ)}_\alpha$, $\bar \Psi^{(IJ)}_{\dot \alpha}$
with $\Delta = 5/2$. The highest-weight components of $\Psi$, $\bar \Psi$
are Schur operators and yield elements of the chiral algebra
with holomorphic dimension $3/2$,
\beq
\mathsf G = \chi[\Psi^{11}_+] \ , \qquad 
\tilde {\mathsf G} = \chi[\bar \Psi^{11}_{\dot +}] \ .
\eeq
The operators $\mathsf G$, $\tilde {\mathsf G}$ are supersymmetry
currents in two-dimensions. 
Both $\Psi$, $\bar \Psi$ and $\mathsf G$, $\mathsf G$ carry implicitly
flavor symmetry indices associated to the commutant of the 
$\cN =2$ R-symmetry $\mathfrak{sl}(2)_R \oplus \mathbb C_r$
inside the larger R-symmetry group of the $\cN = 3$ or $\cN = 4$ theory.

Focusing on the case of an $\cN = 4$ theory,  
the larger (complexified) R-symmetry group is $\mathfrak {sl}(4)_R$
and the commutant of $\mathfrak{sl}(2)_R \oplus \mathbb C_r$ is $\mathfrak{sl}(2)_F$. 
The relevant branching rule is
\beq
\begin{array}{ccc}
\mathfrak{sl}(4)_R & \longrightarrow & \mathfrak{sl}(2)_R \times \mathfrak{sl}(2)_F \times \mathbb C_r \\[2mm]
[1,0,0] & = & \left(\tfrac 12  , 0  \right)_{1/2} \oplus
\left(0 , \tfrac 12 \right)_{-1/2} \ ,
\end{array}
\eeq
where we denoted the fundamental representation of $\mathfrak{sl}(4)_R$  
by its Dynkin indices, $\mathfrak{sl}(2)$ representations by their
half-integral spin,
and the subscript is the $\mathbb C_r$ charge. 
Fundamental indices of $\mathfrak{sl}(2)_F$ will be denoted
$\hat I ,  \hat J = 1,2$.
It follows that the chiral algebra always contains
the two-dimensional small $\cN = 4$ chiral algebra \cite{Ademollo:1976pp}.
The latter is generated by the stress tensor $\mathsf T$,
two supersymmetry currents $\mathsf G^{\hat I}$, $\tilde {\mathsf G}^{\hat I}$
with holomorphic dimension $3/2$ in the fundamental
of $\mathfrak{sl}(2)_F$, and an $\mathfrak{sl}(2)_F$ current $\mathsf J^{(\hat I \hat J)}$.
The Virasoro modes $\mathsf L_{0,\pm1}$,
  the supercurrent modes $\mathsf G^{\hat I}_{\pm 1/2}$,
$\tilde {\mathsf G}^{\hat I}_{\pm 1/2}$ 
and the modes $\mathsf J_0^{\hat I \hat J}$ of the 
affine current
generate a global $\mathfrak {psl}(2|2)$ symmetry.



\section{Localization argument}
\label{sec:localization}

Given a four-dimensional $\cN= 2$ theory admitting a holographic
 dual, it is natural to ask what is the bulk analog of the 
field-theoretic cohomological construction that we have just reviewed.
In this section we address  this problem in a simplified model.

The superconformal algebra on the field theory side
is realized on the gravity side as the algebra of superisometries
of the background. In particular, the background admits suitable
Killing spinors that can be identified with the linear combinations
$\qq_{\,1}$, $ \qq_{\,2}$ of section \ref{sec:review}.
In light of the cohomological construction on the field theory side,
we expect the following picture on the gravity side. 
If we only switch on sources dual to twisted translated Schur
operators on the field theory side,
the partition function on the gravity side should 
be subject to supersymmetric localization and should
define an effective dynamics localized on an $AdS_3$
slice of the original $AdS_5$ spacetime.
The boundary of the $AdS_3$ slice is identified 
  to the preferred $(\zeta, \bar \zeta)$ plane
singled out by the cohomological construction
on the field theory side.  
Implementing this program rigorously  appears challenging in any realistic holographic duality,
{\it e.g.}, in the canonical duality between large $N$ ${\cal N}=4$ SYM  theory and IIB string theory on $AdS_5 \times S^5$.
We are not aware of the requisite  off-shell formalism for IIB supergravity on $AdS_5 \times S^5$, or even for its consistent truncation to ${\cal N}=8$ gauged
supergravity on $AdS_5$.
We can, however, address explicitly a simplified version of the problem,
along the following lines.

Consider an $\cN =2 $ SCFT  with a flavor symmetry  algebra $\mathfrak g_F$. Our main target is ${\cal N}=4$ SYM theory,  for which $\mathfrak g_F = \su(2)_F$ (the centralizer
of the $4d$ ${\cal N}=2$ superconformal algebra $\su(2, 2 |2)$  inside the ${\cal N}=4$ superconformal algebra  ${\mathfrak {psu}}(2, 2 |4)$), but  we may as well keep $\mathfrak g_F$ general.
According to the standard AdS/CFT dictionary, on the gravity side we 
find massless gauge fields with gauge algebra $\mathfrak g_F$, which must belong
to an ${\cal N}=4$ vector multiplet (half-maximal susy). The vector multiplet 
is part of the spectrum of a suitable half-maximal supergravity in five dimensions 
admitting an $AdS_5$ vacuum. We will consider the truncation of the full supergravity to 
the ${\cal N}=4$
 supersymmetric five-dimensional gauge theory  with gauge algebra $\mathfrak g_F$
on a non-dynamical $AdS_5$ background. 
This setup {\it can} be explicitly analyzed using available
localization techniques.

We should point out from the outset that the 
 restriction to the vector multiplet is not  a \emph{bona fide} consistent truncation
 of the full equations of motion of five-dimensional
 supergravity.\footnote{
This can be understood from the viewpoint of  the boundary CFT.
The $5d$ gauge field $A_\mu$ is dual to a conserved current $J_m$, of conformal dimension three. The  {\it singular} OPE of two currents contains  the stress tensor $T_{m n}$,  of conformal dimension four. This is reflected in the bulk in the presence of a cubic coupling between two gauge fields and the fluctuation of the metric $h_{\mu \nu}$, which cannot be removed
by a field redefinition. Setting $h_{\mu \nu} =0$ is not a consistent truncation of the equations of motion, because the equation of motion for $h_{\mu\nu}$ would still induce a spurious constraint for the gauge fields.
By contrast, in the chiral algebra the Sugawara
stress tensor appears (by definition) as the leading {\it non-singular} term in the OPE of two affine currents.} It is however guaranteed to be a ``twisted" consistent truncation,
{\it i.e.}, to hold  in $\qq$-cohomology. Indeed,  the corresponding sector of the chiral algebra is  just the affine Kac-Moody algebra $\widehat g_F$, which is
clearly a closed subalgebra.

\subsection{Summary of the localization results} 

As the details of our calculations are somewhat technical, we begin with a summary of
the main results.
Our goal is to show that the
five-dimensional super Yang-Mills action defined on 
$AdS_5$ localizes to an effective action defined on an
$AdS_3$ slice inside $AdS_5$,
and determine this effective action.
The relevant $AdS_3$ slice is specified as follows.
We can write the Euclidean  $AdS_5$ background in Poincar\'e
coordinates as
\beq
ds^2_5 = \frac{R^2}{z^2} \Big[
d\zeta d\bar \zeta + d\rho^2 + \rho^2 d \varphi + dz^2
\Big] \ ,
\eeq
where $R$ is the $AdS_5$ radius, $z$ is the $AdS_5$ radial coordinate,
$\zeta, \bar \zeta$ are   complex coordinates on a selected plane
on the boundary, $\rho, \varphi$ are polar coordinates 
along the two other directions on the boundary.
The coordinates $\zeta, \bar \zeta$ are identified 
with those 
  used in section \ref{sec:review}
in the discussion of the chiral algebra.
In particular, the plane selected by the cohomological
construction is the plane  spanned by 
$\zeta, \bar \zeta$. 
With this notation, the relevant $AdS_3$ slice
of $AdS_5$ is the one located at $\rho = 0$ and
spanned by $\zeta, \bar \zeta, z$,
\beq
ds^2_3 = \frac{R^2}{z^2} \Big[ d\zeta d\bar \zeta + dz^2 \Big] \ .
\eeq

Let us remind the reader that the bosonic field content of
maximal super Yang-Mills theory in five dimensions
consists of a gauge connection $A$ and five real adjoint scalars,
denoted here $\phi_6$, $\phi_7$, $\phi_8$, $\phi_9$, $\phi_0$.
(Our terminology is related to the ten-dimensional origin of these
fields, described in the following subsection.)
The realization of off-shell supersymmetry used in the localization
computation induces a split
of the five scalars into $(\phi_6, \phi_7)$ and $(\phi_8, \phi_9, \phi_0)$.

After these preliminaries, we can exhibit 
the value of the localized super Yang-Mills action.
It can be written as the sum of two decoupled contributions,
\beq \label{summary_classical_action_as_sum}
S_{} = S_\text{free} + S_{\rm CS}  \ ,
\eeq
where
\begin{align}
S_\text{free} & =  \frac{i \pi R}{g_{\rm YM}^2} \int_{AdS_3} d\zeta d\bar \zeta dz \, 
\frac{R^2}{z^2} \, {\rm tr}\, (\phi_6^2 + \phi_7^2) \ , \\
S_{\rm CS} &= - \frac{ik}{4\pi} \int_{AdS_3} {\rm tr} \left(
\mathbf A  \, d \mathbf A + \frac 23 \mathbf A^3
 \right) \ , \qquad 
 k = - \frac{8\pi^2R}{g_{\rm YM}^2}
 \ .
\end{align}
Here $g_{\rm YM}^2$ denotes the Yang-Mills coupling of the
five-dimensional super Yang-Mills theory,
and the symbol ${\rm tr}$ stands for the trace
in a reference representation of the gauge algebra
(the fundamental for gauge algebra $\mathfrak {su}(N)$).
The scalars $\phi_6$, $\phi_7$   are implicitly evaluated 
at $\rho = 0$, {\it {\it i.e.}}~on the $AdS_3$ slice of $AdS_5$.
The object $\mathbf A$ is an emergent
complex  gauge connection
living on the $AdS_3$ slice of $AdS_5$. 
Its expression in terms of the fields of the original
Yang-Mills theory reads
\beq
\mathbf A = \mathbf A_\zeta \, d\zeta + 
\mathbf A_{\bar \zeta} \, d\bar \zeta
+ \mathbf A_z \, dz \ ,
\eeq
with
\begin{align}
\mathbf A_\zeta &=  A_\zeta - \frac{iR^2}{2z^2} \left[  (\phi_8 + i\, \phi_9)
+ \frac{2 i \bar \zeta}{R} \phi_0 
- \frac{\bar \zeta^2}{R^2} (\phi_8 - i \, \phi_9)
\right] \ , \\
\mathbf A_{\bar \zeta} &=  A_{\bar \zeta} - \frac i2  (\phi_8 - i \, \phi_9) \ , \\
\mathbf A_z &=  A_z + \frac{R}{z}
\left[ 
\phi_0  + \frac{i \bar \zeta}{R} (\phi_8 - i \, \phi_9)
\right]   \ .
\end{align}
The symbols $A_\zeta$, $A_{\bar \zeta}$, $A_z$ denote the components
of the pullback of the original Yang-Mills connection $A$
from $AdS_5$ to the $AdS_3$ slice. The scalars 
$\phi_8$, $\phi_9$, $\phi_0$
are  implicitly evaluated on the $AdS_3$ slice.

The quadratic action $S_\text{free}$ for $\phi_{6,7}$
is expected to be completely decoupled from the rest of the dynamics
on the $AdS_3$ slice, even if suitable supersymmetric insertions
are considered in the path integral. As a result,
$\phi_{6,7}$ are expected to provide only an inconsequential
field-independent Gaussian factor in the computation of 
correlators, and can be effectively ignored.
The emergent gauge field $\mathbf A$, on the other hand,
has dynamics specified by the Chern-Simons action $S_{\rm CS}$, which according to the classic results of \cite{Witten:1988hf, Moore:1989yh,Elitzur:1989nr} defines
 a WZWN theory 
on the boundary of $AdS_3$ based on the group $G_F$. This
 provides a realization of the two-dimensional 
affine current algebra of the Lie algebra $\mathfrak g_F$, as expected from
the cohomological construction on the field theory side.

The outline of the rest of this section is as follows.
The derivation of the above results is described in subsections
\ref{sec:susy_Lagrangian}, \ref{sec:choice_of_susy},
\ref{sec:classical_action}. Further comments about our results
are collected in subsection \ref{sec:remarks},
in which we also test our findings against 
predictions from the chiral algebra based on the 
case of $\cN = 4$ super Yang-Mills wth gauge algebra
$\mathfrak{su}(N)$.

\subsection{Lagrangian with off-shell supersymmetry} \label{sec:susy_Lagrangian}

We consider maximally supersymmetric Yang-Mills theory in five dimensions
on a Euclidean $AdS_5$ background.
Following \cite{Pestun:2007rz, Minahan:2015jta} 
this theory can be constructed in two steps. 
Firstly, the flat-space $10d$ maximally supersymmetric Yang-Mills theory with 
 signature $(1,9)$  is formally dimensionally reduced on a  five-torus with signature
$(1,4)$. Secondly, the external flat metric is replaced with the curved 
Euclidean $AdS_5$ metric, minimal coupling to gravity is introduced,
as well as extra non-minimal couplings needed for supersymmetry.
In order to set up our notation, we review here the field content,
Lagrangian, and off-shell supersymmetry variations,
following closely \cite{Minahan:2015jta}.

Curved $5d$ spacetime indices are denoted $\mu, \nu = 1, \dots 5$.
We adopt Poincar\'e coordinates for the background $AdS_5$ metric,
\beq \label{AdSmetric}
ds^2 = g_{\mu\nu} dx^\mu dx^\nu =
 \frac{R^2}{z^2} \left[ (dx^1)^2 + \dots +(dx^4)^2 + dz^2
 \right] \ , \qquad
x^5 \equiv z \ .
\eeq
A convenient choice of vielbein is
\beq \label{vielbein}
e^{\hat \lambda} {}_\mu = \frac{R}{z} \, \delta^\lambda{}_\mu \ , 
\qquad \lambda,\mu = 1,\dots ,5 \ ,
\eeq
where a hat is used to denote flat $5d$ spacetime indices.

All dynamical bosonic fields of the $5d$ theory originate from the $10d$
 gauge connection
$A_M$, $M = 0, 1, \dots,9$, where $0$ denotes the time direction.
Upon dimensional reduction we obtain the $5d$ gauge connection $A_\mu$,
$\mu = 1,\dots ,5$, as well as five real scalars $\phi_\cI \equiv A_\cI$, 
$\cI = 6,\dots,9,0$ in the adjoint representation of the gauge group.
The index $\cI$ is a vector index of the R-symmetry
 $\mathfrak {so}(4,1)_R$. The latter, however, 
  is   explicitly
broken to $\mathfrak{so}(2)_R \oplus \mathfrak{so}(2,1)_R$
by the way 
off-shell supersymmetry is realized   below.
Correspondingly, it is useful to introduce  the notation
\beq
\phi_\cI = (\phi_i , \phi_A) \ , \qquad i = 6,7 \ , \quad A = 8,9,0 \ .
\eeq
We use anti-Hermitian generators for the gauge algebra
and
the $10d$ field strength reads
\beq
F_{MN} = 2 \partial_{[M} A_{N]} + [A_M , A_N]  \ .
\eeq
Its components after dimensional reduction are given by
\begin{align}
F_{\mu\nu} &=  2 \partial_{[\mu} A_{\nu]} + [A_\mu , A_\nu] \ , & 
F_{\mu \cI} &= \partial_\mu \phi_\cI   + [A_\mu , \phi_\cI] \equiv  D_\mu \phi_\cI  \ ,& 
F_{\cI \cJ} &=  [\phi_\cI  , \phi_\cJ]  \ .
\end{align}
In order to close supersymmetry off-shell we also need to introduce
seven real auxiliary scalars $K^m$, $m = 1, \dots, 7$ in the adjoint
representation of the gauge group. Their vector $\mathfrak{so}(7)$ index
is raised and lowered with the flat invariant $\delta_{mn}$.

 All fermionic degrees of freedom are encoded in a 16-component
Grassmann-odd
$10d$ Majorana-Weyl gaugino $\Psi_\alpha$, $\alpha = 1, \dots, 16$,
in the adjoint representation of the gauge group.
The chiral blocks of $10d$ gamma matrices are denoted
$\Gamma^M{}^{\alpha \beta}$,  $\tilde \Gamma^M{}_{\alpha \beta}$,  
and we also use the notation
\beq
\Gamma^{MN} = \tilde \Gamma^{[M} \Gamma^{N]}  \ , \quad 
\tilde \Gamma^{MN} =   \Gamma^{[M} \tilde  \Gamma^{N]}  \ , \quad
\Gamma^{MNP} = \Gamma^{[M} \tilde \Gamma^N \Gamma^{P]} \ , \quad
\tilde \Gamma^{MNP} = \tilde  \Gamma^{[M}   \Gamma^N  \tilde \Gamma^{P]}  \ .
\eeq
Weyl indices are henceforth suppressed. After dimensional reduction
and coupling to the curved $AdS_5$ background, the $10d$ covariant derivative 
of the gaugino $D_M \Psi$ gives rise in five dimensions to
\beq
D_\mu \Psi = \partial_\mu \Psi 
+ \tfrac 14 \omega_{\mu \hat \lambda \hat \tau} \Gamma^{\hat \lambda \hat \tau}
\Psi + [A_\mu , \Psi] , \qquad D_\cI \Psi = [\phi_\cI, \Psi] \ ,
\eeq
where
$\omega_{\mu \hat \lambda \hat \tau}$ is the spin connection associated to the 
background vielbein \eqref{vielbein}.

The off-shell supersymmetric Lagrangian reads
\begin{align} \label{5dLagr}
\cL = \frac{1}{g_{\rm YM}^2} \, {\rm tr} \bigg[& \frac 12 F_{MN} F^{MN}
- \Psi \Gamma^M D_M \Psi
+ \frac {i}{2R} \, \Psi \Lambda \Psi
- \frac{3}{R^2} \phi^i \phi_i
- \frac{4}{R^2} \phi^A \phi_A \nn \\
&
- \frac{2i}{3R} \epsilon^{ABC}[\phi_A , \phi_B]\phi_C
- K^m K_m 
\bigg] \ ,
\end{align}
where
${\rm tr}$ denotes the trace in a reference representation
(the fundamental for gauge algebra $\mathfrak{su}(N)$), and 
 we defined
\beq \label{def_Lambda}
\Lambda = \Gamma^8 \tilde \Gamma^9 \Gamma^0 \ , \qquad
\epsilon_{890}  = +1 \ , \qquad
\epsilon^{890} = -1 \ .
\eeq
Note that we adopted the customary compact notation
\beq \label{explicitFMNFMN}
F_{MN} F^{MN} = F_{\mu\nu} F^{\mu\nu} + 2 D_\mu \phi_\cI D^\mu \phi^\cI 
+ [\phi_\cI , \phi_\cJ] [\phi^\cI , \phi^\cJ] \ ,
\eeq
in which the spacetime indices $\mu,\nu$ are curved and thus
raised with the metric \eqref{AdSmetric}, while the indices
$\cI$, $\cJ$ are flat and raised with the 
 $\mathfrak {so}(4,1)_R$ metric
 $\eta_{\cI \cJ} = {\rm diag} (1,1,1,1,-1)$.
In a similar way we have
\beq \label{explicitDslashPsi}
\Gamma^M D_M \Psi = \Gamma^\mu D_\mu \Psi  + \Gamma^\cI [\phi_\cI , \Psi] 
= \Gamma^{\hat \lambda}  \, e_{\hat \lambda} {}^\mu D_\mu \Psi  + \Gamma^\cI [\phi_\cI , \Psi]   \ ,
\eeq
where $e_{\hat \lambda} {}^\mu$ denotes the inverse of the $5d$ vielbein
\eqref{vielbein}. 
All spinor bilinears in \eqref{5dLagr} and in the following
are Majorana bilinears. Further
details about our spinor conventions are collected in appendix 
\ref{app:conventions}.

The Lagrangian \eqref{5dLagr} is invariant up to total derivatives
under the off-shell supersymmetry transformations
\begin{align}
\delta A_\mu &= \epsilon \Gamma_\mu \Psi \ ,   \label{offshell_susy1}   \\
\delta \phi_\cI & =\epsilon \Gamma_\cI \Psi \ , \label{offshell_susy2}   \\
\delta \Psi & = \frac 12 \Gamma^{MN} F_{MN} \epsilon 
+ \frac 25 \Gamma^{\mu i} \phi_i \nabla_\mu \epsilon
+ \frac 45 \Gamma^{\mu A} \phi_A \nabla_\mu \epsilon 
+ K^m \nu_m\ , \label{offshell_susy3}   \\
\delta K^m & = - \nu^m \Gamma^M D_M \Psi + \frac{i}{2R} \nu^m \Lambda \Psi \ .
\label{offshell_susy4} 
\end{align}
In these expressions $\epsilon$ is a Grassmann-even 16-component
Majorana-Way spinor with the same chirality as $\Psi$.
It satisfies the $AdS_5$ Killing spinor equation
\beq \label{KSE}
\nabla_\mu \epsilon =\frac{i}{2R} \tilde \Gamma_\mu \Lambda   \epsilon  \ .
\eeq
Note that in this equation
$\tilde \Gamma_\mu = \tilde \Gamma_{\hat \lambda} e^{\hat \lambda}{}_\mu$,
$\nabla_\mu \epsilon = \partial_\mu \epsilon + \tfrac 14 \omega_{\mu \hat \lambda
\hat \tau} \Gamma^{\hat \lambda \hat \tau} \epsilon$.
Let us also stress that the compact notation $\Gamma^{MN} F_{MN}$
in \eqref{offshell_susy3}
is subject to remarks similar to those around \eqref{explicitFMNFMN} 
and \eqref{explicitDslashPsi} above. 
We have also introduced a set $\nu_m$, $m = 1,\dots, 7$ of  auxiliary Grassmann-even
spinors with the same chirality as $\epsilon$,
determined up to an $\mathfrak{so}(7)$ rotation by the algebraic relations
\beq \label{pure_spinor_eqs}
\nu_m \Gamma^M \epsilon = 0 \ , \qquad
\nu_m \Gamma^M \nu_n = \delta_{mn} \, \epsilon \Gamma^M \epsilon \ .
\eeq
The $\mathfrak {so}(7)$ index $m$ on $\nu_m$ is raised with $\delta^{mn}$.

The square of the supersymmetry transformations 
\eqref{offshell_susy1}-\eqref{offshell_susy4}
can be written as combination of the bosonic symmetries of the theory
without using the equations of motion. More precisely, one has
\begin{align}
\delta^2 A_\mu & = - v^\nu F_{\nu\mu} + D_\mu (v^\cI \phi_\cI)  \ ,
\label{susy_squared1} \\
\delta^2 \phi_i & =  - v^\nu D_\nu \phi_i - [v^\cI \phi_\cI  ,  \phi_i ]
- \frac i R \, \epsilon \tilde \Gamma_{ij} \Lambda \epsilon \, \phi^j \ , 
\label{susy_squared2}\\
\delta^2 \phi_A & =  - v^\nu D_\nu \phi_A - [v^\cI \phi_\cI  ,  \phi_A ]
- \frac {2i}{ R} \, \epsilon \tilde \Gamma_{AB} \Lambda \epsilon \, \phi^B \ ,
\label{susy_squared3} \\
\delta^2\Psi & = - v^\nu D_\nu \Psi - \frac 14 \nabla_\mu v_\nu \, \Gamma^{\mu\nu} \Psi
 - [v^\cI \phi_\cI, \Psi] 
  - \frac{i}{2R} (\epsilon \tilde \Gamma^{AB} \Lambda \epsilon) \Gamma_{AB} \Psi
- \frac{i}{4R} (\epsilon \tilde \Gamma^{ij} \Lambda \epsilon) \Gamma_{ij} \Psi \ ,
\label{susy_squared4}
 \\
\delta^2 K^m & =  - v^\nu D_\nu K^m 
 - [v^\cI \phi_\cI, K^m]
+ \left(
-  \nu^{[m} \Gamma^\mu \nabla_\mu \nu^{n]} 
+ \frac{i}{2R} \nu^{[m} \Lambda \nu^{n]}  \right) K_n  \ ,
\label{susy_squared5}
\end{align}
where we utilized the spinor bilinears
\beq \label{vector_bilinear}
v^\mu = \epsilon \Gamma^\mu \epsilon = e{}_{\hat \lambda}{}^\mu \,
  \epsilon \Gamma^{\hat \lambda} \epsilon       \ , \qquad
v^\cI = \epsilon \Gamma^\cI \epsilon \ .
\eeq
The $5d$ vector $v^\mu$ is a Killing vector for the $AdS_5$
background metric, $\nabla_{(\mu} v_{\nu)}   = 0$.
Note that $\delta^2 K^m$ contains an $\mathfrak{so}(7)$ rotation,
which is a symmetry of the Lagrangian.

All our formulae can be obtained as an analytic continuation
of the formulae given in \cite{Minahan:2015jta} for the case of the five-sphere.
More precisely, the radius $r$ of $S^5$ is related to the radius $R$ of Euclidean
$AdS_5$ as
\beq
R = i r \ . 
\eeq
Note, however, that the coordinate system utilized in \cite{Minahan:2015jta}
is different from the one adopted here, and would correspond in the case of
Euclidean $AdS_5$ to the disk model of hyperbolic space, rather than
the half-space model.

\subsection{Identification of the relevant  supercharge}   \label{sec:choice_of_susy}

Our first task in the implementation of the localization argument
is the identification of the Killing spinor corresponding to the relevant
supercharge on the field theory side.
As reviewed in section \ref{sec:review}, for a unitary theory
$\qq_{\,1}$ and $\qq_{\,2}$ define the same cohomology classes
on the field theory side. From the point of view of localization
it is most convenient to consider 
\beq
\qq = \qq_{\, 1} + \qq_{\, 2} \ .
\eeq
The holomoprhic
$\mathfrak{sl}(2)$ factor on the fixed plane is $\qq$-closed,
and the twisted antiholomorphic factor 
$\widehat{\mathfrak{sl}(2)}$ is $\qq$-exact. 
Note, however, that $\qq$ is not nilpotent,
but rather satisfies
\beq
\{ \qq \, , \qq\,  \} = 2 \{ \qq_{\, 1} , \qq_{\, 2} \} = -2 (r + \cM_\perp) \ .
\eeq
On the gravity side, if we select the Killing spinor $\epsilon$
corresponding to $\qq\,$, the associated Killing vector $v$
contains the spacetime action $\cM_\perp$,
consisting of rotations in the directions orthogonal
to the fixed plane.
As a result, we localize on the fixed point set of $\cM_\perp$,
consisting of the fixed plane itself.

In order to identify the Killing spinor
corresponding to $\qq$ 
we   have to analyze the space of solutions to the
Killing spinor equation \eqref{KSE} with $\Lambda$
given in \eqref{def_Lambda}. We refer the reader to 
  section \ref{app:Killing_spinors} in the appendices
for a thorough discussion and for the 
explicit expression for the Killing spinor~$\epsilon$.
 Let us summarize here some of its key properties. 
To this end, it is convenient to use complex
coordinates $\zeta$, $\bar \zeta$ in the $x^1x^2$ plane
and polar coordinates $\rho$, $\varphi$ in the $x^3x^4$ plane,
\beq
\zeta = x^1 + i x^2 \ , \qquad 
\bar \zeta = x^1 - i x^2 \ , \qquad
x^3 = \rho \, \cos \varphi \ , \qquad
x^4  = \rho \, \sin \varphi \ .
\eeq 
The Killing vector $v^\mu$ defined in \eqref{vector_bilinear} takes the form
\beq \label{KillingVector}
v^\mu \partial_\mu = \partial_\phi \ ,
\eeq
while the field-dependent gauge parameter that enters the square of the
supersymmetry transformations \eqref{susy_squared1}-\eqref{susy_squared5}
 is given by
\beq \label{our_gauge_parameter}
v^\cI \phi_\cI =\frac{i R   \rho}{z} (\cos \varphi \, \phi_6 - \sin \varphi \, \phi_7) \ .
\eeq
Our Killing spinor induces no $\mathfrak{so}(2,1)_R$ R-symmetry rotation,
but yields a non-zero $\mathfrak{so}(2)_R$ rotation,
\beq
\epsilon \tilde \Gamma^{AB} \Lambda \epsilon = 0 \ , \qquad
 \epsilon \tilde \Gamma^{ij} \Lambda \epsilon =i\, R\,  \epsilon^{ij} \ , \qquad
\epsilon^{67} = +1  \ .
\eeq
Recall that the Lie derivative of a spinor in the direction of a Killing
vector $k^\mu$ is given by
\beq
\cL_k \epsilon = k^\mu \nabla_\mu \epsilon
+ \frac 14 \nabla_\mu k_\nu \, \Gamma^{\mu\nu} \epsilon \ .
\eeq
Using this expression one can check that our Killing spinor is
invariant under the action of the Killing vectors associated to 
the holomorphic conformal generators in the $(\zeta, \bar \zeta)$ plane.
More precisely,
if we consider the Killing vectors
\begin{align}
k(L_{-1}) & = -   \partial_\zeta   \ , \\
k(L_0) & = - \zeta \partial_\zeta - \tfrac 12 \rho \partial_\rho - \tfrac 12 z \partial_z \ , \\
k(L_{+1}) & = - \zeta^2 \partial_\zeta 
- \zeta \rho \partial_\rho - \zeta z \partial_z
+ (z^2 + \rho^2) \partial_{\bar \zeta} \ ,
\end{align}
we have
\beq
\cL_{k(L_m)} \epsilon = 0 \ , \qquad m = -1,0,+1 \ .
\eeq
This corresponds to the fact that our supercharge commutes with the 
holomorphic conformal generators  in the $(\zeta, \bar \zeta)$ plane.
Furthermore, we expect the anti-holomorphic generators to be exact.
This expectation is confirmed by checking that each of the Killing vectors
\begin{align}
k(\overline L_{-1}) & = -   \partial_{\bar \zeta}   \ , \\
k(\overline L_0) & = - \bar \zeta \partial_{\bar \zeta}
 - \tfrac 12 \rho \partial_\rho - \tfrac 12 z \partial_z \ , \\
k(\overline L_{+1}) & = - \bar \zeta^2 \partial_{\bar \zeta} 
- \bar \zeta \rho \partial_\rho - \bar \zeta z \partial_z
+ (z^2 + \rho^2) \partial_{  \zeta} \ ,
\end{align}
can be written in the form $\
epsilon \Gamma^\mu \epsilon'$ for a suitable
Killing spinor $\epsilon'$.

Once the suitable Killing spinor $\epsilon$ is identified,
we are left with the task of finding the associated auxiliary
spinors $\nu_m$ satisfying \eqref{pure_spinor_eqs}.
We refer the reader to section \ref{app:auxiliary} in the appendix
for more details on this point. 

\subsection{BPS locus and classical action} \label{sec:classical_action}
  
The localization argument ensures that in the computation of $\qq$-closed
observables the path integral localizes to the BPS locus
\beq
\Psi  = 0 \ , \qquad \delta \Psi = 0 \ .
\eeq
In particular this implies that $\delta^2$ annihilates all fields on the BPS locus.  
Making use of the expression for $\delta^2$ recorded in the previous
section one can verify that, for our choice of supercharge, this implies  
\beq
D_{\varphi} \phi_\cI   = -  [v^\cJ \phi_\cJ , \phi_\cI] \ , \qquad
F_{\rho \varphi}   = - D_\rho (v^\cI \phi_\cI)  \  , \qquad
F_{\tilde \mu \varphi}   = - D_{\tilde \mu} (v^\cI \phi_\cI)  \ ,
\eeq
where we introduced a new $3d$ curved spacetime index $\tilde \mu = \zeta, \bar \zeta,z$
and   $v^\cI \phi_\cI$ is given in \eqref{our_gauge_parameter}.
Let us point out that, in all the above equations, 
the covariant derivative acts on spacetime scalars and therefore
contains the gauge field but no spacetime connection.

Once the constraints coming from $\delta^2 = 0$ are implemented,
one can show that the 16 equations $\delta \Psi  = 0$ 
are all solved by determining the seven auxiliary scalars $K^m$ as a functional
of all other bosonic fields. 
In summary,
\beq \label{BPSlocus}
\text{BPS locus:} \quad
\left\{ \;\;\;\;
\begin{array}{l}
\Psi = 0 \ , \\[2mm]
D_{\varphi} \phi_\cI   = -  [v^\cJ \phi_\cJ , \phi_\cI] \ , \quad
F_{\rho \varphi}   = - D_\rho (v^\cI \phi_\cI)  \  , \quad
F_{\tilde \mu \varphi}   = - D_{\tilde \mu} (v^\cI \phi_\cI) \ , \\[2mm]
\text{$K^m$ determined in terms of $A_\mu$, $\phi_\cI$.}
\end{array}
\right.
\eeq    
We refrain from recording here the explicit expressions for the auxiliary scalars $K^m$ 
in terms of $A_\mu$, $\phi_\cI$, which are lengthy and not particularly illuminating.

As a next step in the localization we 
evaluate the classical Lagrangian \eqref{5dLagr} on the BPS locus
\eqref{BPSlocus}.  
A straightforward but tedious computation shows that
the entire bosonic Lagrangian, including the appropriate
volume form, collapses on the BPS locus to a sum of total
derivatives.
More precisely, one finds
\beq \label{L_total_der}
\sqrt g \, \cL \, d^4x \, dz = \left[ 
\partial_\rho Y^\rho + \partial_\zeta Y^\zeta
+ \partial_{\bar \zeta} Y^{\bar \zeta}
+ \partial_{\varphi} Y^{\varphi}
+ \partial_z Y^z
\right]
d\zeta d \bar \zeta d \rho d\varphi dz \ ,
\eeq
where the quantities $Y$ are suitable functionals of the gauge
field and scalars whose explicit expressions are not 
recorded for the sake of brevity.
On the LHS the notation $d^4 x dz$ is a shorthand for
the five form
$dx^1 \wedge dx^2 \wedge dx^3 \wedge dx^4 \wedge dz$,
and by a similar token we have omitted
wedge products on the RHS. 
In checking \eqref{L_total_der} it is essential to take into account the 
factors coming from the expression of the $AdS_5$ volume
form in the $(\zeta, \bar \zeta, \rho , \varphi, z)$ coordinate system,
\beq
\sqrt {g} \, d^4x \, dz  = \frac{i R^5 \rho}{2 z^5} \, 
d\zeta   d \bar \zeta   d \rho   d\varphi      dz\ .
\eeq
The classical action on the BPS locus is given by the integral of 
\eqref{L_total_der} over the factorized domain
\beq
- \infty < {\rm Re} \zeta < \infty \ , \quad
- \infty < {\rm Im} \zeta < \infty \ , \quad
0 \le \rho < \infty \ , \quad
0 \le \varphi < 2\pi \ , \quad
0 \le z < \infty \ .
\eeq
Let us discuss the possible boundary contributions.
Of course, since all fields are periodic in the angular variable $\varphi$
no boundary term can be generated by integrating $\partial_\varphi Y^\varphi$.
We assume that all fields fall off sufficiently rapidly 
at infinity in  all directions orthogonal to the radial coordinate $z$ of $AdS_5$.
As a result, we get no contributions from $\partial_\zeta Y^\zeta
 + \partial_{\bar \zeta} Y^{\bar \zeta}$, while $\partial_\rho Y^\rho$
 contributes exclusively via the lower limit of integration $\rho = 0$.
The asymptotic behavior of fields in the $z$ direction
is more subtle and is related to the implementation
of the AdS/CFT prescription for the computation of correlators.
The goal of our localization computation is the identification
of an effective $3d$ bulk theory that could  be then used to 
compute correlators of twisted-translated Schur operators
according to the standard prescription.
For the purpose of identifying the $3d$ theory we do not need to
consider boundary terms coming from the $z$ direction.

In conclusion, the relevant classical action on the BPS locus
can be written as
\beq \label{implicit_classical_action}
S_{\rm cl} =2\pi \int d\zeta d\bar \zeta   dz\left( - Y^\rho \big|_{\rho = 0} \right) \ ,
\eeq
where we anticipated that $Y^\rho$ is actually independent of $\varphi$
and we performed the $\varphi$ integration.
The fact that all fields are evaluated at $\rho = 0$
shows manifestly the expected localization of the dynamics
on the $\zeta\bar \zeta$ plane which is fixed under the action
of the Killing vector \eqref{KillingVector} associated to 
our Killing spinor.

Let us now  record the expression of the integrand in 
\eqref{implicit_classical_action}
in a convenient way.
To this end, it is useful to trade the scalar fields 
$\phi_A$, $A = 6,7,8$ of the original Yang-Mills theory
with the components of a one-form $\Phi_{\tilde \mu}$
living on the $AdS_3$ slice of $AdS_5$ identified by $\rho = 0$
and parametrized
by $\zeta$, $\bar \zeta$, $z$.
This twist is achieved by means of a suitable object $\mathscr V^{A\tilde \mu} $
built from bilinears of the Killing spinor $\epsilon$ in the following way.
To begin with, let us define the $5d$ three-vector\beq
\mathscr X^A {}^{\mu_1 \mu_2 \mu_3} =-\frac{i}{2R}  \,  \epsilon \Lambda  \tilde \Gamma^A \Gamma^{\mu_1 \mu_2 
\mu_3} \, \epsilon \ ,
\eeq
where $\mu_1$, $\mu_2$, $\mu_3$ are curved $5d$ indices.
Our choice of spinor breaks $5d$ covariance by selecting
the plane spanned by $\rho$, $\varphi$. 
It is thus natural to consider the components
$\mathscr X^A {}^{\tilde \mu  \rho \varphi}$
with $\tilde \mu =\zeta, \bar\zeta , z$.
The sought-for intertwiner $\mathscr V^{A\tilde \mu}$
is then constructed as
\beq
\mathscr V^{A \tilde \mu} = \lim_{\rho \rightarrow 0^+} \rho \, 
\mathscr X^{A  \tilde \mu \rho \varphi}  \ ,
\eeq
where the prefactor $\rho$ has been introduced to guarantee finiteness of the limit.
The relation between the scalars $\phi_A$ and the twisted one-form $\Phi_\mu$
is then
\beq
\phi_A = \frac{2 R^2}{z^2} \,\Phi_{\tilde \mu} \, \eta_{AB} \mathscr V^{B \tilde \mu} \ , \qquad
\eta_{AB} = {\rm diag}(1,1,-1) \ ,
\eeq
where  the normalization factor
has been  chosen for later convenience,
and all fields are implicitly evaluated at $\rho = 0$.
More explicitly, in our conventions the components of $\Phi_{\tilde \mu}$
are given by
\begin{align}   
\Phi_\zeta & =   - \frac{iR^2}{2z^2} \left[  (\phi_8 + i\, \phi_9)
+ \frac{2 i \bar \zeta}{R} \phi_0 
- \frac{\bar \zeta^2}{R^2} (\phi_8 - i \, \phi_9)
\right] \ , \label{explicit_Phi1}\\
\Phi_{\bar \zeta} & =  - \frac i2  (\phi_8 - i \, \phi_9) \ ,
\label{explicit_Phi2} \\
\Phi_z & =   \frac{R}{z}
\left[ 
\phi_0  + \frac{i \bar \zeta}{R} (\phi_8 - i \, \phi_9)
\right] \ . \label{explicit_Phi3}
\end{align}

We can finally present the explicit expression for the
classical action \eqref{implicit_classical_action}.
It can be written   as the sum of
a non-topological and a topological term,
\beq \label{classical_action_as_sum}
S_{\rm cl} = S_\text{free} + S_{\rm CS}  \ ,
\eeq
where
\begin{align}
S_\text{free} & =  \frac{i \pi R}{g_{\rm YM}^2} \int_{AdS_3} d\zeta d\bar \zeta dz \, 
\frac{R^2}{z^2} \, {\rm tr}\, \phi^i \phi_i \ , \\
S_{\rm CS} & =  \frac{2 \pi i R}{g_{\rm YM}^2} \int_{AdS_3} {\rm tr} \left( \Phi \, d_A \Phi 
+ \frac 23 \Phi^3
+ 2 F \, \Phi
+ A \, dA + \frac 23 A^3
\right) \label{S-top}  \ .
\end{align}
Let us remind the reader that all quantities are implicitly evaluated at $\rho = 0$.
In the second line  we have adopted a differential
form notation suppressing wedge products and,
by slight abuse of notation, $A$, $F$ denote the 
restriction of the $5d$ gauge connection and field strength the 
$AdS_3$ slice spanned by coordinates $\zeta$, $\bar \zeta$, $z$.
The symbol $d_A$ denotes the exterior gauge-covariant derivative
\beq
d_A \Phi = d\Phi + A \Phi + \Phi A \ .
\eeq
Let us point out that the appearance of the topological
term ${\rm tr}\, \Phi^3$ in \eqref{S-top} is a   consequence of the cubic term
$\epsilon^{ABC} \phi_A [\phi_B , \phi_C]$
in the scalar potential of the original
Yang-Mills Lagrangian \eqref{5dLagr}.

It is useful to construct the quantity
\beq \label{def_mathbfA}
\mathbf A = A + \Phi \ ,
\eeq
which transforms as a connection
since $\Phi$ is an adjoint-valued one-form. 
Thanks to the identity
\begin{align}
{\rm tr} \left (  \mathbf A \, d\mathbf A + \frac 23 \mathbf A^3\right ) 
& = {\rm tr}  \left(  A dA + \frac 23 A^3 + \Phi d_A \Phi + \frac 23 \Phi^3
+ 2 \Phi  F \right) + d {\rm tr} (\Phi A) \ ,
\end{align}
the topological part of the action $S_{\rm top}$ can be written 
compactly as a Chern-Simons action,
\beq \label{Stop_expression}
S_{\rm CS} = - \frac{ik}{4\pi} \int_{AdS_3} {\rm tr} \left(
\mathbf A  \, d \mathbf A + \frac 23 \mathbf A^3
 \right) \ , \qquad 
 k = - \frac{8\pi^2R}{g_{\rm YM}^2} \ .
\eeq 
The minus sign in front of the Chern-Simons term is
introduced because, in our conventions, the pairing
${\rm tr}(ab)$ is negative definite, since we are using
antihermitian generators. For istance, for gauge algebra $\mathfrak{su}(n)$\footnote{This
$5d$/$3d$ gauge algebra should not be confused with the gauge algebra
$\mathfrak{su}(N)$ of $4d$ $\cN = 4$ SYM. The case $n=2$
 will be relevant below, where
we make contact to the chiral algebra dual to $4d$ $\mathfrak{su}(N)$ $\cN = 4$ SYM
by specializing the $5d$ gauge algebra to be $\mathfrak{su}(2)_F \supset \mathfrak{su}(4)_R$, the R-symmetry algebra of SYM.}
we adopt the standard normalization  
with ${\rm tr}$ denoting the trace in the fundamental representation,
\beq
{\rm tr}(t_\mathbf a \, t_\mathbf b) = - \tfrac 12 \delta _{\mathbf a \mathbf b} \ ,
\qquad
[t_\mathbf a , t_\mathbf b] = f_{\mathbf a \mathbf b}{}^{\mathbf c} \, t_{\mathbf c} \ ,
\eeq
where $\mathbf a , \mathbf b = 1, \dots , n^2-1$ are adjoint indices
of $\mathfrak{su}(n)$. As a result, we may also write
\beq \label{Stop_expressionBIS}
S_{\rm CS} =  \frac{ik}{8\pi} \, \delta_{\mathbf a \mathbf b} \int_{AdS_3}
\left( 
\mathbf A^\mathbf a \, d\mathbf A^\mathbf b
+ \frac 13 f_{\mathbf c \mathbf d}{}^{\mathbf b}  \mathbf A^\mathbf a 
\mathbf A^\mathbf c \mathbf A^\mathbf d
\right)
 \ , \qquad 
 k = - \frac{8\pi^2R}{g_{\rm YM}^2} \ .
\eeq

The localization technique can be applied
with arbitrary insertions of $\qq$-closed functionals of the fields.
The functionals should be well-defined on the  $AdS_3$ slice and
have a  vanishing supersymmetry variation 
 \eqref{offshell_susy1}-\eqref{offshell_susy2} on that slice.
 It is not hard to check that functionals 
built  from 
$\mathbf A_{\zeta}$, $\mathbf A_{\bar \zeta}$ and
$\mathbf A_z$ (and independent of $\phi_{6,7}$) satisfy these requirements,\footnote{It also appears that these are the {\it only} admissible functionals.
At first sight,
the
linear combinations
\beq
   A_\rho + \frac{R}{2z} \left[ 
e^{i \varphi} (\phi_6 + i \phi_7)
- e^{- i \varphi} (\phi_6 - i \phi_7)
\right] \ , \quad
   A_\varphi \ ,
\eeq
seem admissible, since they have a
 vanishing supersymmetry variation for $\rho \rightarrow 0$. However,
assuming that the Cartesian components $A_3$, $A_4$ of $A_\mu$
are smooth near $\rho = 0$, 
the quantity $A_\rho = \cos \varphi \, A_3 + \sin \varphi \, A_4$
does not admit a unique  limit for $\rho \rightarrow 0$,
but rather depends on the $\varphi$ angle with which we approach
$\rho =0$, while the quantity
$A_\varphi = \rho(- \sin \varphi \, A_3 + \cos \varphi \, A_4)$
vanishes for $\rho \rightarrow 0$.
These combinations must therefore be discarded.
} 
in agreement with the 
conclusion that 
$\phi_{6,7}$ can be consistently decoupled.
 
\subsection{Remarks} \label{sec:remarks}

Our implementation of the localization recipe is different from the 
one usually applied to supersymmetric theories on Euclidean
compact manifolds. In the latter case it is customary to supplement
the classical Lagrangian by an explicit   $\qq$-exact localizing term, 
$S_{\rm tot} = S + t\, \qq V$. The functional $V$ and the reality conditions
on the fields are chosen in such a way as to guarantee that,
as $t \rightarrow \infty$, the path integral 
converges and localizes on a suitable real slice of (a subspace of)
the BPS locus.   Different choices for  $V$ and for the reality conditions yield 
different localization schemes. 
The computation of the previous section shows that,
in {\it any} localization scheme, the classical action must 
reduce to \eqref{classical_action_as_sum}.
This conclusion only relies on the form of BPS locus
\eqref{BPSlocus} without choosing a specific real 
slice in the space of field configurations. 
For example, the BPS locus allows for a non-zero
profile for the scalars $\phi_i$, but they enter the
classical action via the quadratic, algebraic action
$S_{\text{non-top}}$ only, and therefore decouple from the 
dynamics after localization.

We are ready to go back to our main physical goal --
 the determination of the bulk
holographic dual to the $2d$ chiral algebra. The
 bulk action \eqref{classical_action_as_sum} must be
supplmented with suitable
boundary conditions for the fields
$\mathbf A$, $\phi_{6,7}$ in order to implement the holographic recipe
for the computation of correlators in the boundary theory.
We should also contemplate the possibility of additional boundary terms
to the $AdS_3$ action. The boundary conditions and boundary terms
for the 
theory on the $AdS_3$ slice could be derived 
via localization of an appropriate set of boundary conditions
and boundary terms
in the original super Yang-Mills theory defined in $AdS_5$.
These $5d$ data are constrained by the requirement of compatibility
with the action of the supercharge $\qq$ selected for
localization.\footnote{We 
refer the reader to \cite{DiPietro:2015zia,Assel:2016pgi} for a discussion of these points in
similar contexts.}
For the problem at hand, we can follow a simpler route
without making reference to the parent $5d$ bulk theory.
Since we have already argued that the scalars $\phi_{6,7}$
play no relevant role, we focus on $\mathbf A$ only.

To begin with, we observe that supercharge $\qq$
induces an asymmetry in the treatment of holomorphic
and antiholomorphic components $\mathbf A_\zeta$, $\mathbf A_{\bar \zeta}$.
This is most easily detected by looking at the 
 the expressions
\eqref{explicit_Phi1}-\eqref{explicit_Phi3} for the components of
the twisted one-form $\Phi$. 
Inspection of \eqref{explicit_Phi1}-\eqref{explicit_Phi3} reveals
a hierarchy of the three components with respect to the
radial coordinate of $AdS_3$. In particular,
if we prescribe the boundary conditions
$\phi_{8,9,0} \sim z^2$ as $z \rightarrow 0$
 in order to get a finite $\Phi_\zeta$,
 then  $\Phi_{\bar \zeta}$
 and $\Phi_{z}$ necessarily vanish at the boundary.
We can regard $\mathbf A_\zeta$, $\mathbf A_{\bar \zeta}$,
$\mathbf A_z$ as the supersymmetrizations
of $\Phi_\zeta$,  $\Phi_{\bar \zeta}$, $\Phi_{z}$,
and argue that the asymmetric pattern for the   holomorphic 
and antiholomorphic components persists.
This mechanism is the bulk dual of the
emergence of a purely meromorphic
dynamic on the field theory side,
once we consider cohomology classes
of the supercharges $\qq_1$, $\qq_2$.

From this observation we can deduce the correct  boundary terms that must be added to the action.
As explained for instance in \cite{Elitzur:1989nr,Gukov:2004id,Kraus:2006nb,Kraus:2006wn},  meromorphic boundary 
conditions are selected by supplementing the
 Chern-Simons action with
the boundary term
\beq \label{bdy_action}
S_{\rm bdy} =   \frac{s\, k}{8\pi} \int_{\partial AdS_{3}} {\rm tr} (\mathbf A 
* \mathbf A)
=- \frac{s\, k}{16\pi} \, \delta_{\mathbf a \mathbf b} 
 \int_{\partial AdS_{3}} d^2 x \sqrt g \, g^{\mu\nu} 
 \mathbf A^{\mathbf a}_\mu \,
 \mathbf A^{\mathbf b}_\nu \ ,
\eeq
where $s$ is a constant to be fixed momentarily
and, by slight abuse of notation, we used $\mu,\nu$
to denote two dimensional curved  indices on $\partial AdS_3$.
The combined variation of the bulk action \eqref{Stop_expression}
and the boundary action \eqref{bdy_action}
with respect to the gauge field takes the form
\beq
\delta (S_{\rm CS} +   S_{\rm bdy}) = 
- \frac{ik}{2\pi} \int_{AdS_3} {\rm tr}(\delta \mathbf A \,
\mathbf F)
- \frac{ik}{2\pi} \int_{\partial AdS_3} d\zeta d\bar \zeta \,
 {\rm tr}( \delta \mathbf A_\zeta  \, \mathbf J_{\bar \zeta}
 + \delta \mathbf A_{\bar \zeta}  \, \mathbf J_{  \zeta}) 
  \ ,
\eeq
where
\beq
\mathbf J_{\bar \zeta} = \tfrac 12 (1-s) \mathbf A_{\bar \zeta} \ , \qquad
\mathbf J_{  \zeta} = \tfrac 12 (1+s) \mathbf A_{  \zeta} \ .
\eeq
The bulk term in the variation imposes that the connection
be flat. The  currents $\mathbf J_{  \zeta}$,
$\mathbf J_{\bar  \zeta}$ entering the boundary terms of the variation
are identified with the currents of the boundary CFT.
Because of our choice of supercharge 
we know that the antiholomorphic component
$\mathbf J_{\bar  \zeta}$ of the boundary current
should be zero. Thus, we must select $s = 1$. This implies that $k_{\rm 2d} = k$, which is negative in our case (recall (\ref{Stop_expressionBIS})), in agreement with field theory expectations.

Our choice  differs from the one  in \cite{Kraus:2006wn},  where $s = {\rm sgn}\, k$ is advocated on the basis of the following argument.
The boundary action \eqref{bdy_action} contributes to the 
boundary stress tensor,
\beq
\delta S_{\rm bdy} =\tfrac 12 \int_{\partial AdS_3} d^2 x \sqrt g \,
 T^{\mu\nu}_{\rm bdy} \,
\delta g_{\mu\nu} \ , \qquad
T^{\mu\nu}_{\rm bdy} = \frac{s\, k}{8 \pi} \, \delta_{\mathbf a \mathbf b}
\Big[ 
\mathbf A^{\mathbf a \, \mu} \,\mathbf A^{\mathbf b \, \nu}
- \frac 12 g^{\mu\nu} \, \mathbf A^{\mathbf a \, \tau} \,\mathbf A^{\mathbf b}_\tau
\Big] \ , 
\eeq
which in complex components reads
\beq
T^{\rm bdy}_{\zeta \zeta} = \frac{s\, k}{8\pi} \, \delta_{\mathbf a \mathbf b}
\, \mathbf A^{\mathbf a}_\zeta
\, \mathbf A^{\mathbf b}_{  \zeta} \ , \qquad
T^{\rm bdy}_{\bar \zeta  \bar \zeta} =
 \frac{s\, k}{8\pi} \, \delta_{\mathbf a \mathbf b}
\, \mathbf A^{\mathbf a}_{\bar \zeta}
\, \mathbf A^{\mathbf b}_{\bar \zeta} \ , \qquad
T^{\rm bdy}_{\zeta \bar \zeta} = 0 \ .
\eeq
As explained in \cite{Kraus:2006wn}, with these conventions
 a positive coefficient of $\mathbf A^{\mathbf a}_\zeta
\, \mathbf A^{\mathbf b}_{  \zeta}$ in  $T^{\rm bdy}_{\zeta \zeta}$
 corresponds to a
positive definite contribution
to the boundary energy in a semi-classical picture, 
leading to the  prescription $s = {\rm sgn} \, k$ and $k_{\rm 2d} = |k|$. 
In other terms, in the standard case unitarity of the boundary CFT is enforced by hand. In our case, 
we must let supersymmetry dictate the correct boundary conditions, and we naturally land on a non-unitary chiral algebra.

We can now specialize \eqref{Stop_expression} to the case in which the 
boundary theory is $4d$ $\cN = 4$ super Yang-Mills with gauge algebra $\mathfrak{su}(N)$.
Regarded as a $4d$ $\cN = 2$ theory, this theory
has a flavor symmetry $\mathfrak{su}(2)_F$, the commutant of
$\mathfrak{su}(2)_R \times \mathfrak{u}(1)_r$ inside $\mathfrak {su}(4)_R$.
The gravity dual of $\cN =4$ super
Yang-Mills  is maximal gauged supergravity with gauge group
$\mathfrak{su}(4)_R$.
The corresponding gauge coupling function,
evaluated at the origin of the scalar potential
corresponding to  the $AdS_5$ vacuum, is
\cite{Freedman:1998tz}
\beq
\frac {g_{\mathfrak{su}(4)_R}^2 }{R} = \frac{16 \pi^2}{N^2} \ ,
\eeq
at leading order in $1/N$. 
We have to consider the branching
$\mathfrak{su}(4)_R \rightarrow \mathfrak{su}(2)_R \times \mathfrak{u}(1)_r \times \mathfrak{su}(2)_F$
and restrict to the $\mathfrak{su}(2)_F$ factor.
One can easily check that this does not affect the normalization of the Yang-Mills
kinetic term, so  $g_{\rm YM} =g_{\mathfrak{ su}(2)_F}= g_{\mathfrak{ su}(4)_R}$.
As a result, in this case the Chern-Simons level $k$ in \eqref{Stop_expression}
is
\beq
k =-  \tfrac 12 N^2 \ .
\eeq
We can compare this result with the level of the affine current
algebra of the $\mathfrak{su}(2)_F$ current $\mathsf J^{\hat I \hat J}$ in the
chiral algebra dual to $\cN =4$ super Yang-Mills
with gauge algebra $\mathfrak{su}(N)$. One finds \cite{Beem:2013sza}
\beq
k_{\rm 2d} = - \tfrac 12  k_{\rm 4d} = - \tfrac 12 (N^2-1) \ , 
\eeq
which agrees with
 the Chern-Simons level $k$  \eqref{Stop_expression} in the large $N$ limit.

 Finally, 
let us briefly comment about  quantum corrections to the classical
result \eqref{classical_action_as_sum}. 
While it may be of some technical interest to compute
the one-loop  determinant factor 
associated to
fluctuations of  super Yang-Mills fields transversely to the 
localization locus,  the physical relevance of such a calculation is {\it a priori} unclear. What would be physically relevant is a calculation of quantum fluctuations in a fully consistent holographic theory,
{\it e.g.}~in  IIB string field theory on $AdS_5 \times S^5$, but this is clearly beyond the scope of this work -- indeed even the classical problem seems prohibitively hard in the complete theory.
Given the agreement of  \eqref{classical_action_as_sum} with the expected large $N$ result, we may speculate that the only effect of quantum fluctuations is an $O(1)$ shift of the Chern-Simons level.


\section{Towards the complete holographic dual}

In this section we propose a strategy to determine the complete holographic dual for the chiral algebra of ${\cal N}=4$ SYM theory.
A straightforward task is the identification of the linearized bulk modes that correspond to non-trivial boundary operators in $\qq$-cohomology.
Their non-linear interactions are however extremely complicated,  and extending  the localization procedure to the full theory is presently
beyond our technical abilities. Encouraged by the emergence of a Chern-Simons action in the simplified model discussed above, and drawing inspiration
from minimal model holography, we will outline a bottom-up construction of the dual theory. We will argue that it is a Chern-Simons theory
with gauge algebra a suitable higher-spin Lie superalgebra, defined implicitly by the large $N$ OPE coefficients of ${\cal N}=4$ SYM.

We begin in section 4.1 with a review of the super chiral algebra conjecture of \cite{Beem:2013sza}. 
We give  a simple argument in favor of the conjecture in the large $N$ limit. The generators\footnote{A {\it generator} of a chiral algebra is an operator that does {\it not} appear in the non-singular
OPE of other operators. In the mathematics literature, what we call a generator is usually referred to as a {\it strong generator}.}
 of the chiral algebra are 
the single-trace Schur operators, which are in 1-1 correspondence with KK sugra modes obeying the Schur condition.
In section 4.2 we give the details of this correspondence. Following the blueprint of minimal model holography,
we propose in section 4.3 that the sought after holographic dual  is $AdS_3$ Chern-Simons theory with gauge algebra
given by the  wedge algebra of the large $N$ super W-algebra. While it is unclear whether such a wedge algebra  exists for finite $N$, we 
outline its construction at infinite $N$.

\subsection{The  $\cN = 4$ SYM chiral algebra at large $N$}
\label{sec:SYM_chiral_review}

 As we have reviewed in section 2.2, the chiral algebra 
 always admits  the small ${\cal N}=4$ superconformal algebra  (SCA) as a subalgebra. 
  The global part 
 of the    small ${\cal N}=4$ SCA is $\mathfrak {psu}(1, 1|2)$, and we henceforth organize the operator content in terms of $\mathfrak {psu}(1, 1|2)$ primaries and their descendants.
   It was conjectured in \cite{Beem:2013sza} that the chiral algebra  associated to ${\cal N}=4$ SYM 
with gauge algebra $\su(N)$ is  generated by $N-1$ 
$\mathfrak {psu}(1,1|2)$  primaries, obeying the shortening condition $h = k$, where $h$ is the holomorphic dimension ($L_0$ eigenvalue) and $k$ the $\su(2)_F$ spin.
These supergenerators have $h=k =\tfrac 12 (n +2)$
where $n = 0, \dots, N-2$, and will be denoted as 
$\mathsf J^{(n)}{}^{\hat I_1 \dots \hat I_{n+2}}$.
The central charge is fixed by the general formula (\ref{c2d_vs_c4d}) to $c_{\rm 2d } = - 3 (N^2 -1)$. 

The $\mathfrak {psu}(1,1|2)$ primary operator 
$\mathsf J^{(n)}{}^{\hat I_1 \dots \hat I_{n+2}}$ is clearly also an
$\mathfrak {sl}(2)$ primary, and 
among its superdescendants 
we find additional $\mathfrak {sl}(2)$ primary
operators, which we denote as
$\mathsf G^{(n)}{}^{   \hat I_1 \dots \hat I_{n+1}}$,
$\tilde{ \mathsf G}^{(n)}{}^{   \hat I_1 \dots \hat I_{n+1}}$,
$\mathsf T^{(n)}{}^{\hat I_1 \dots \hat I_n}$.\footnote{A
 fully explicit statement of the super W-algebra conjecture
 for $N\ge 3$
is then the following: (i) as a vector space, the chiral algebra
is the linear span of derivatives of 
$\mathsf J^{(n)} $,
$\mathsf G^{(n)} $,
$\tilde{ \mathsf G}^{(n)} $,
$\mathsf T^{(n)} $ ($n = 0,\dots , N-2$)
and their conformally ordered products; (ii)
the operators 
$\mathsf J^{(n)} $,
$\mathsf G^{(n)} $,
$\tilde{ \mathsf G}^{(n)} $,
$\mathsf T^{(n)} $ ($n = 0,\dots , N-2$)
cannot be written as conformally ordered products
of derivatives of other operators.
}
The schematic form of the supermultiplet
containing $\mathsf J^{(n)}{}^{\hat I_1 \dots \hat I_{n+2}}$,
$\mathsf G^{(n)}{}^{   \hat I_1 \dots \hat I_{n+1}}$,
$\tilde{ \mathsf G}^{(n)}{}^{   \hat I_1 \dots \hat I_{n+1}}$,
$\mathsf T^{(n)}{}^{\hat I_1 \dots \hat I_n}$ is
\beq  \label{psu112_pattern}
\begin{array}{l}
\vspace{-1.75mm}  \\  h =1 + \tfrac 12  n   \vspace{0.65cm }  \\ \vspace{0.65cm } 
h =\tfrac 32  + \tfrac 12  n  \\ 
 h =2 + \tfrac 12  n 
\end{array} \quad \quad \quad
\begin{tikzcd}[row sep = scriptsize]
{}  				& \mathsf J^{(n)}{}^{\hat I_1 \dots \hat I_{n+2}}    \drar {{ \tilde Q}_{\hat I}}  \dlar[swap]{{Q}_{\hat I}}	& {} \\
\mathsf G^{(n)}{}^{\hat I_1 \dots \hat I_{n+1}} 
\drar[swap]{ {\tilde Q}_{\hat I} } 	&  	
&
\tilde {\mathsf G}^{(n)}{}^{\hat I_1 \dots \hat I_{n+1}} 
  \dlar{Q_{\hat I}} \\
{}					
& \mathsf T^{(n)}{}^{\hat I_1 \dots \hat I_{n}}	& {}
\end{tikzcd}  \ ,
\eeq
where $Q_{\hat I}$, $\tilde Q_{\hat I}$ denote   the  Poincar\'e 
supercharges of $\mathfrak{psu}(1,1|2)$.  For gauge algebra $\su(2)$, the conjecture asserts that there is a unique supergenerator, the $\su(2)_F$ affine current $\mathsf J^{(0)}{}^{\hat I \hat J} \equiv \mathsf J^{\hat I \hat J}$.
The full $\mathfrak {psu}(1, 1|2)$ supermodule comprises $\mathsf J^{\hat I \hat J}$,   the supercurrents $\mathsf G^{(0)}{}^{\hat I} \equiv \mathsf G^{\hat I}$ and $\tilde {\mathsf G}^{(0)}{}^{\hat I} \equiv \tilde {\mathsf G}^{\hat I}$ and the stress tensor 
$\mathsf T^{(0)} \equiv \mathsf T$, which is of course the operator content of the small ${\cal N}=4$ SCA.\footnote{
In fact, for $c_{\rm 2d } = - 9$,  the stress tensor $\mathsf T$ is not
an independent generator, but is rather identified
with the Sugawara stress tensor built from the affine current
$\mathsf J^{\hat I \hat J}$.} In this special case there is clearly no obstruction in deforming the central charge to arbitrary values. 

The  supergenerators   $\mathsf J^{(n)}$
in 1-to-1 correspondence with the generators of the $1/2$ BPS chiral ring of the SYM theory, {\it i.e.}~with the single-trace  $1/2$ BPS  operators  of the
  form  ${\rm tr}\, Z^{n+2}$, with $n = 0, \dots, N-2$
(see \eqref{XZZbar_def} below for the  definition of $Z$).
In terms of $\mathfrak{psu}(2, 2 |4)$ representation theory, these operators are the bottom components of 
 multiplets of type 
$\cB^{ \scriptscriptstyle{ \frac 12 ,\frac  12}}_{[0,n+2,0](0,0)}$
in the notations of \cite{Dolan:2002zh}.
An $\mathfrak{psu}(2, 2 |4)$ multiplet of type 
$\cB^{ \scriptscriptstyle{ \frac 12 ,\frac  12}}_{[0,n+2,0](0,0)} $
decomposes into $\mathfrak{su}(2, 2 |2)$
as \cite{Dolan:2002zh}
\begin{align} \label{rep_decomposition}
\cB^{ \scriptscriptstyle{ \frac 12 ,\frac  12}}_{[0,n+2,0](0,0)} & = 
(n+3) \hat \cB_{{\scriptscriptstyle \frac 12} (n+2)}
\oplus (n+2) \left[ 
\cD_{{\scriptscriptstyle \frac 12 }(n+1)(0,0)}
\oplus
\bar \cD_{{\scriptscriptstyle \frac 12 }(n+1)(0,0)}
\right]
\oplus
(n+1) \hat \cC_{{\scriptscriptstyle \frac 12 }n(0,0)}  
\oplus \dots 
\\
& \hspace{2.cm } \Big \downarrow
\hspace{3.1cm} \Big \downarrow
\hspace{2cm} \Big \downarrow
\hspace{3cm} \Big \downarrow \nn
\\
&\hspace{1.3cm } \mathsf J^{(n)}{}^{\hat I_1 \dots \hat I_{n+2}}
\hspace{1.55cm} \mathsf G^{(n)}{}^{   \hat I_1 \dots \hat I_{n+1}}
\hspace{0.4cm} \tilde{ \mathsf G}^{(n)}{}^{   \hat I_1 \dots \hat I_{n+1}}
\hspace{1.6cm} \mathsf T^{(n)}{}^{\hat I_1 \dots \hat I_n}  \nn
\end{align}
Below each shown  $\mathfrak{su}(2, 2 |2)$ multiplet 
 we have indicated the
corresponding chiral algebra operator, which arises from 
as the $\qq$-cohomology class of the Schur operator in the multiplet.
(The ellipsis in \eqref{rep_decomposition}
represents  additional  multiplets
 that are not relevant for
our discussion,
since they do not    contain Schur operators.)
 The degeneracies in \eqref{rep_decomposition}
are  accounted for by the dimensions
of the $\mathfrak {su}(2)_F$ representations.
 
The fact that the operators $\mathsf J^{(n)}$
are  supergenerators of the  chiral algebra
is easily established. Indeed, they arise from $4d$ 1/2 BPS operators, which are absolutely protected against quantum corrections;\footnote{Recall that $1/2$ BPS  multiplets  {\it cannot} recombine into long multiplets \cite{Dolan:2002zh}.} as they correspond to generators of the 1/2 BPS chiral ring it is clear that they  {\it cannot}  appear in the non-singular OPE of other operators.
The hard  part of the conjecture of \cite{Beem:2013sza} is showing that these are {\it all} the supergenerators. The chiral algebra is specified by a non-trivial BRST procedure, which in physical terms
amounts to selecting  operators obeying the Schur shortening condition in the interacting theory.

The conjecture, however, can be proved at infinite $N$. 
All Schur operators are in particular $1/16$ BPS operators of 
$\cN = 4$ SYM, {\it i.e.}, operators in the cohomology of a single Poincar\'e supercharge.
This cohomology was studied in \cite{Chang:2013fba},
where it was proved that at infinite $N$ it is obtained by taking arbitrary 
 products of $1/16$ BPS   single-trace operators,\footnote{The  product operation relevant here is the commutative product induced by the ordinary OPE of the $4d$ theory, which is non-singular
for $1/16$ BPS operators.} which were further shown to be in 1-1 correspondence with single 1/16 BPS gravitons in the dual $AdS_5 \times S^5$ supergravity.
Schur operators are in the simultaneous cohomology of {\it two} Poincar\'e supercharges of opposite chirality, say $\cQ^1_-$ and $\tilde \cQ_{2 \dot -}$ in the conventions of \cite{Beem:2013sza}.
Specializing the results of  \cite{Chang:2013fba} to this double cohomology, we find that at infinite $N$ it is given  by arbitrary products\footnote{In this statement, the product operation is again
 the commutative product induced by the standard OPE of the $4d$ theory. The twisted translation prescription of \cite{Beem:2013sza} deforms this commutative algebra into the chiral algebra that we are interested in.
There is an easy general argument that the  generators of the chiral algebra are a subset of the generators of the commutative algebra. This is just what we need (the other inclusion is obvious in our case).} of the single-trace Schur operators corresponding to  $\mathsf J^{(n)}$,  $\mathsf G^{(n)}$,   $\mathsf {\tilde G}^{(n)}$,   $\mathsf T^{(n)}$, $n \in \mathbb{Z}_+$. This shows that these operators comprise
the full set of  generators for the chiral algebra at infinite $N$.

\subsection{Single-trace Schur operators of $\cN = 4$ SYM and supergravity}
\label{sec:Schur_and_KK}

Having established the super chiral algebra conjecture of \cite{Beem:2013sza}
for infinite $N$, we now proceed to
give more details on the single-trace Schur operators 
in \eqref{rep_decomposition} and to map them to the 
Kaluza-Klein modes in type IIB supergravity on $AdS_5 \times S^5$ \cite{Kim:1985ez}.
Our conclusions are summarized in table \ref{table:Schur}.

\begin{table}
\centering
\begin{tabular}{r||c|c|c|cc}
supermultiplet
&
$\hat \cB_{{\scriptscriptstyle \frac 12} (n+2)}$
& 
$\cD_{{\scriptscriptstyle \frac 12 }(n+1)(0,0)} $
& 
$\bar \cD_{{\scriptscriptstyle \frac 12 }(n+1)(0,0)}$
&
$(n+1) \hat \cC_{{\scriptscriptstyle \frac 12 }n(0,0)}$
&
\parbox{1mm}{\vspace{0.8cm}}
\\
superprimary
&
${\rm tr}\, ( Q^{n+2})$
& 
${\rm tr}\, ( Z\,  Q^{n+1})$
& 
${\rm tr}\, ( \bar Z\,  Q^{n+1})$
&
${\rm tr}\, ( Z\, \bar Z \,  Q^{n})$
&
\parbox{1mm}{\vspace{0.8cm}}
\\
Schur operator
&
${\rm tr}\, ( Q^{n+2})$
& 
${\rm tr}\, ( \lambda_+ \, Q^{n+1})$
& 
${\rm tr}\, ( \tilde \lambda_{\dot +} \, Q^{n+1})$
&
${\rm tr}\, ( \lambda_+ \, \tilde \lambda_{\dot +} \, Q^{n})$
&
\parbox{1mm}{\vspace{0.8cm}}
\\
$\Delta$
&
$n+2$
& 
$n + \tfrac 52$
& 
$n + \tfrac 52$
&
$n+3$
&
\parbox{1mm}{\vspace{0.8cm}}
\\
chiral algebra operator
&
$\mathsf J^{(n)}{}^{\hat I_1 \dots \hat I_{n+2}}$
& 
$\mathsf G^{(n)}{}^{\hat I_1 \dots \hat I_{n+1}}$
& 
$\tilde {\mathsf G} ^{(n)}{}^{\hat I_1 \dots \hat I_{n+1}}$
&
$\mathsf T^{(n)}{}^{\hat I_1 \dots \hat I_{n}}$
&
\parbox{1mm}{\vspace{0.8cm}}
\\
$h$
&
$\tfrac 12 (n+2)$
& 
$\tfrac 12 (n+3)$
& 
$\tfrac 12 (n+3)$
&
$\tfrac 12 (n+4)$
&
\parbox{1mm}{\vspace{0.8cm}}
\\
$J_F$
&
$\tfrac 12 (n+2)$
& 
$\tfrac 12 (n+1)$
& 
$\tfrac 12 (n+1)$
&
$\tfrac 12 n$
&
\parbox{1mm}{\vspace{0.8cm}}
\\
KK mode
&
\parbox[b][2pt][b]{2.5cm}{\centering $\pi^{I_1}$}
& 
\parbox[b][2pt][b]{2.5cm}{\centering $\psi^{I_L}$}
& 
\parbox[b][2pt][b]{2.5cm}{\centering $\psi^{I_L}$}
&
\parbox[b][2pt][b]{2.5cm}{\centering $B_\mu^{I_5}$}
&
\parbox{1mm}{\vspace{0.8cm}}
\\
KK mass
&
$m^2 R^2  =n^2-4 $
& 
$|m|  R  =n + \tfrac 12 $
& 
$|m|  R  =n + \tfrac 12 $
&
$|m|  R  =n^2+2n $
&
\parbox{1mm}{\vspace{0.8cm}}
\end{tabular}
\caption{Families of Schur operators of $\cN = 4$ super Yang-Mills theory.
For each family we give the $\cN = 2$ supermultiplet
in the notation of \cite{Dolan:2002zh},
the schematic form of the superprimary in the multiplet,
and the schematic form of the  Schur operator.
The quantum numbers $\Delta$, $h$, $J_F$ are the
$4d$ scaling dimension of the Schur operator,
the $2d$
holomorphic dimension of the chiral
algebra element,
and the half-integer $\mathfrak{sl}(2)_F$
spin of both, respectively.
The scalar fields $Q$, $Z$, $\bar Z$ are defined in
\eqref{XZZbar_def}, while 
$\lambda$, $\tilde \lambda$ denote the gaugini
in $\cN = 2$
language,
which are a subset of all gaugini in $\cN =4$
language.
The KK modes are given in the 
notation of \cite{Kim:1985ez}.  
All families are labelled in such a way that the range of $n$
is $n = 0,1,2, \dots$.
}
\end{table} \label{table:Schur}

All Kaluza-Klein modes in the compactification of type IIB supergravity
on $AdS_5 \times S^5$ are dual to operators 
that are organized in $1/2$-BPS short $\cN = 4$ multiplets of type 
$\cB^{ \scriptscriptstyle{ \frac 12 ,\frac  12}}_{[0,n+2,0](0,0)}$.  
In Lagrangian language, the $\cN = 4$ superprimaries
of these multiplets can be written as
\beq \label{N4_chiral_primaries}
\cO^{(n)}{}^{\cA_1 \dots \cA_{n+2}} = {\rm tr}(X^{ \{ \cA_1} \dots X^{\cA_{n+2} \}}) \ ,
\qquad n = 0, 1, 2, \dots  \ ,
\eeq
where   $\cA_1$, \dots, $\cA_{n+2}$ are vector indices of $\mathfrak{so}(6)_R$,
$X^\cA$ are the real scalars of $\cN = 4$ super Yang-Mills,
and curly brackets denote the traceless symmetric part.

We have already reviewed the decomposition of
an $\cN = 4$ multiplet of type 
$\cB^{ \scriptscriptstyle{ \frac 12 ,\frac  12}}_{[0,n+2,0](0,0)} $
 into superconformal multiplets of $\cN = 2$,
see \eqref{rep_decomposition}.
In order to elucidate the connection between
the branching rule  \eqref{rep_decomposition} and the 
Lagrangian presentation \eqref{N4_chiral_primaries},
it is convenient 
to reorganize the scalars $X^\cA$ schematically as
\beq \label{XZZbar_def}
Z = X^5 + i X^6 \ , \qquad
\bar Z =  X^5 - i X^6 \ , \qquad
Q^{I \hat J} = X^a \, \sigma_a{}^{ I \hat J} \ ,
\eeq
where $a = 1,\dots,4$, $I =1,2$ is a fundamental index of $\mathfrak{su}(2)_R$,
$\hat J = 1,2$ is a fundamental index of $\mathfrak{su}(2)_F$,
and $\sigma_a{}^{  I \hat J}$ are chiral blocks of $\mathfrak{so}(4)$ gamma matrices.
The scalar $Z$ is the complex scalar in the $\cN = 2$ vector multiplet,
while $Q^{I \hat J}$ are the scalars in the $\cN = 2$ hypermultiplet.
We can now easily identify a Lagrangian realization of the $\cN = 2$ superconformal
primary for each of the multiplets on the RHS of \eqref{rep_decomposition}.
Let us list them together with their  $R$- and $F$-isospins
and $r$ charges,
\beq \label{N2_superprimaries}
\begin{array}{rcllll}
 \hat \cB_{{\scriptscriptstyle \frac 12} (n+2)} 
 &    
 : & 
  \quad  
  {\rm tr} (Q^{(I_1| \hat J_1} \dots
 Q^{I_{n+2}) \hat J_{n+2}})  \ ,&
\quad   
J_R = \tfrac 12 (n+2)  \ ,&
\quad  
J_F =\tfrac 12 (n+2)  \ ,&
\quad   
r = 0 \ ,
\\[1mm]
\cD_{{\scriptscriptstyle \frac 12 }(n+1)(0,0)} 
&     
: &
 \quad  
 {\rm tr} (Z \,
Q^{(I_1| \hat J_1} \dots
 Q^{I_{n+1}) \hat J_{n+1}})  \ ,&
  \quad  
J_R = \tfrac 12 (n+1)  \ ,&
\quad  
J_F =\tfrac 12 (n+1) \ , &
\quad   
r = 1 \ ,
\\[1mm]
 \bar \cD_{{\scriptscriptstyle \frac 12 }(n+1)(0,0)} 
&      
 : &
  \quad  
 {\rm tr} (\bar Z \,
Q^{(I_1| \hat J_1} \dots
 Q^{I_{n+1}) \hat J_{n+1}})  \ ,&
  \quad  
J_R = \tfrac 12 (n+1)  \ ,&
\quad  
J_F =\tfrac 12 (n+1)  \ ,&
\quad   
r = -1 \ ,
\\[1mm]
\hat \cC_{{\scriptscriptstyle \frac 12 }n(0,0)}
 &     
: & 
 \quad  
  {\rm tr} (Z \,\bar Z \,
Q^{(I_1| \hat J_1} \dots
 Q^{I_{n}) \hat J_{n}})  \ ,&
  \quad  
J_R = \tfrac 12 n \ , &
\quad  
J_F =\tfrac 12 n \ , &
\quad   
r = 0 \ .
  \end{array}    \nn
\eeq 
Each of these $\cN = 2$ supermultiplets yields a Schur operator.
Let us discuss them in turn and relate them to the associated
Kaluza-Klein mode in the spectrum of type IIB supergravity on $AdS_5 \times S^5$.

\paragraph{Multiplets of type $ \hat \cB_{{\scriptscriptstyle \frac 12} (n+2)} $.}
In this case the Schur operator is directly the
$\mathfrak{su}(2)_R$  highest-weight component of the superconformal primary
listed in \eqref{N4_chiral_primaries}.
It follows that the operator in the chiral algebra is simply
\beq \label{hatB_Schur_ops}
\mathsf J^{(n)}{}^{\hat I_1 \dots \hat I_{n+2}} = \chi \left[  {\rm tr} (Q^{1 \hat J_1} \dots
 Q^{1 \hat J_{n+2}}) \right] \ , 
\quad
h = \tfrac 12 (n+2) \ , \quad
J_F = \tfrac 12 (n+2) \ , \quad 
n =0,1,\dots \ ,
\eeq
where we summarized its $2d$ quantum numbers.
The gravity duals of the 
$\cN = 4$ chiral primaries in \eqref{N4_chiral_primaries} are given 
by the Kaluza-Klein modes named $\pi^{I_1}$
in Table III of \cite{Kim:1985ez}.
It follows that the gravity duals of the operators 
\eqref{hatB_Schur_ops} are given by the subset of the modes 
$\pi^{I_1}$ 
corresponding to 
the $J_R =\tfrac 12 (n+2)$, $J_F = \tfrac 12 (n+2)$
representation of $\mathfrak {su}(2)_R \oplus  \mathfrak {su}(2)_F$
inside the $[0,n+2,0]$ of $\mathfrak {so}(6)_R$.
The masses of the $\pi^{I_1}$ Kaluza-Klein tower are
\beq
m^2 R^2 =(n-2)(n+2) \ , \qquad
n = 0,1,\dots  
\eeq
The case $n = 0$ deserves special attention. The 
superconformal primary  of $ \hat \cB_{1} $
is the moment map for the $\mathfrak{su}(2)_F$ flavor symmetry.
The associated operator in the chiral algebra
$ \mathsf J^{(0)}{}^{\hat I \hat J} \equiv \mathsf J{}^{\hat I \hat J}$
is the affine $\mathfrak{su}(2)_F$ current of the small $\cN = 4$
subalgebra. The dual scalar mode in supergravity
has a negative mass-squared that saturates 
the Breitenlohner-Friedmann bound
\cite{Breitenlohner:1982jf}.

\paragraph{Multiplets of types $\cD_{{\scriptscriptstyle \frac 12 }(n+1)(0,0)} $
and $ \bar \cD_{{\scriptscriptstyle \frac 12 }(n+1)(0,0)} $.}
In this case the Schur operator is a component of a super-descendant
of the scalar operator listed in 
\eqref{N2_superprimaries}. More precisely,
for $\cD_{{\scriptscriptstyle \frac 12 }(n+1)(0,0)}$
we need to act with $\tilde \cQ^I_{\dot \alpha}$,
obtaining a right-handed spinor operator of the schematic form
\beq \label{Psi_op}
\Psi^{(n)}{}^{I_0 I_1 \dots I_{n+1} \hat I_1 \dots \hat I_{n+1}}_{\dot \alpha}
 = {\rm tr}(\tilde \lambda_{\dot \alpha}^{I_0} \,
Q^{ I_1  \hat J_1} \dots
 Q^{I_{n+1} \hat J_{n+1}})
 + \dots \ ,  \qquad n = 0,1, \dots \ ,
\eeq
where we recorded explicitly the part coming from the 
action of $\tilde \cQ^I_{\dot \alpha}$ on $Z$,
which yields the $\cN =2$ gaugino
$\tilde \lambda_{\dot \alpha}^{I}$,
but we omitted additional terms arising from the action of 
$\tilde \cQ^I_{\dot \alpha}$ on the $\cN = 2$
hypermultiplet scalars.
The quantum numbers of the operator in \eqref{Psi_op}
are
\beq
 J_R = \tfrac 12 (n+2) \ , \qquad
 J_F = \tfrac 12 (n+1) \ , \qquad
 r = \tfrac 12  \ , \qquad
\Delta = n + \tfrac 52 \ ,
\eeq
and its  $\mathfrak{su}(4)_R$ orbit is the one of the $\cN = 4$
superdescendant of \eqref{N4_chiral_primaries}
in the $[0,n+1,1]$ representation of $\mathfrak{su}(4)_R$.

The Schur operator is
the highest-weight component of $\Psi^{(n)}$
and  the associated chiral operator is then
\beq
\mathsf G^{(n)}{}^{   \hat I_1 \dots \hat I_{n+1}}
=\chi \left[ \Psi^{(n)}{}^{1\dots 1  \hat I_1 \dots \hat I_{n+1}}_{\dot +} \right] \ ,
\quad
h = \tfrac 12 (n+3) \ , \quad
J_F= \tfrac 12 (n+1) \ , \quad
n = 0,1, \dots
\eeq
Completely analogous considerations hold for
type  $\bar \cD_{{\scriptscriptstyle \frac 12 }(n+1)(0,0)} $
multiplets. The analog of $\Psi^{(n)}$, denoted 
$\tilde \Psi^{(n)}$,
is built using the supercharge $\cQ_\alpha^I$
and thus contains a $\lambda_\alpha^I$ insertion. It has the same
quantum numbers as $\Psi^{(n)}$, except $r = -\tfrac 12$,
and $\tilde {\mathsf G}^{(n)}$ is the associated  operator in the chiral algebra.

The gravitational dual of the operators 
$\Psi^{(n)}$,   
$\tilde \Psi^{(n)}$
is encoded in the suitable R-symmetry
components of $5d$ Dirac spinor modes denoted 
$\psi^{I_L}$ in Table III of \cite{Kim:1985ez}.
Their masses are
\beq
m R= -  (n + \tfrac 12 ) \ , \qquad 
n = 0, 1, 2, \dots .
\eeq
The minus sign is relative to the positive masses of the
excited
Kaluza-Klein modes in the tower of the $5d$ gravitino.

In the case $n=0$ the multiplets  
$\cD_{{\scriptscriptstyle \frac 12 } (0,0)} $,
 $ \bar \cD_{{\scriptscriptstyle \frac 12 } (0,0)} $ contain
 among their descendants the spin-$3/2$
supersymmetry currents of dimension $7/2$
associated to the supercharges of the $\cN  =4$
superalgebra that are not in the selected $\cN = 2$
subalgebra. The corresponding operators
$\mathsf G^{(0)}{}^{\hat I} \equiv \mathsf G^{\hat I} $
and  $\tilde {\mathsf G}^{(0)}{}^{\hat I} = \tilde {\mathsf G}^{\hat I}$ 
in the chiral algebra
are supersymmetry currents. 

\paragraph{Multiplets of type $\hat \cC_{{\scriptscriptstyle \frac 12 }n(0,0)}$.}
In this case the Schur operator  is a component of the operator
obtained acting with one $\cQ$ and one $\tilde \cQ$ on the
superprimary in \eqref{N2_superprimaries}.
Schematically, we have
\beq
J^{(n)}{}_{\alpha \dot \beta}{}^{K_1 K_2 I_1 \dots I_n
\hat J_1 \dots \hat J_n} = {\rm tr} ( \lambda_\alpha^{K_1} \, \tilde \lambda^{K_2}_{\dot \beta} \,
Q^{I_1 \hat J_1} \dots Q^{I_n \hat J_n}
) + \dots \ ,
\quad n = 0,1,\dots  \ ,
\eeq
where we omitted several other terms for the sake of brevity.
The quantum numbers of this $4d$ operator are
\beq
J_R = \tfrac 12 (n+2) \ , \qquad
J_F = \tfrac 12 n \ , \qquad
r = 0 \ , \qquad
\Delta = n + 3 \ ,
\eeq
and its $\mathfrak{su}(4)_R$ completion is the $\cN = 4$
superdescendant of \eqref{N4_chiral_primaries}
in the $[1,n,1]$ representation of $\mathfrak{su}(4)_R$.
 
The chiral algebra operator is therefore
\beq
\mathsf T^{(n)}{}^{\hat I_1 \dots \hat I_n} = \chi \left[
J^{(n)}{}_{+ \dot +}{}^{1 \dots 1
\hat J_1 \dots \hat J_n} 
\right] \ , \quad
h = \tfrac 12 (n+4) \ , \quad
J_F = \tfrac 12 n \ , \quad
n = 0, 1, \dots
\eeq
The gravity dual to the vector operators $J^{(n)}$ is furnished
by the vector modes
called $B_\mu^{I_5}$ in Table III of \cite{Kim:1985ez},
with masses
\beq
m^2 R^2 =  n (n+2)  \ , \quad n = 0,1,2, \dots 
\eeq
For $n = 0$ the multiplet $\hat \cC_{0(0,0)}$ contains the $4d$ stress tensor
and the Schur operator is a component of the $\mathfrak{su}(2)_R$ symmetry
current. The operator $\mathsf T^{(0)} \equiv \mathsf T$ in the chiral algebra is the $2d$ stress tensor.
On the gravity side, we find the massless vectors associated to the
Killing vectors of $S^5$.

The families discussed  above
have a natural $\mathbb Z_2$ grading corresponding to
even modes $n = 0,2, \dots$, and odd modes $n = 1, 3, \dots$. The 
series for even $n$   constitutes a consistent truncation
of the chiral algebra.  For  $n$ even, $\mathsf J^{(n)}$ and 
$\mathsf T^{(n)}$ have integer holomorphic dimension,
while $\mathsf G^{(n)} $ and $\tilde {\mathsf G}^{(n)}$ 
 have half-integer holomorphic dimension.
These assignments obey the standard spin/statistics connection.
On the other hand, for $n$ odd the situation is reversed, and the spin/statistics connection is violated. There is of course no contradiction -- this is the generic
case for chiral algebras associated to ${\cal N}=2$ SCFTs.

\subsection{Comments on the full higher-spin algebra}

Motivated by the emergence 
of an $AdS_3$ Chern-Simons theory in the localization
computation of section 
\ref{sec:localization}, we believe that  
the bulk
dual of the full chiral algebra is a higher-spin $AdS_3$ Chern-Simons theory.
This expectation is in line with known examples of 
minimal model holography (see \cite{Gaberdiel:2012uj} for a review).
From this perspective, we are left with the task of determining the
correct higher-spin Lie superalgebra in which the Chern-Simons gauge connection
takes values.

Before proceeding, it is useful to review the well-understood
case in which the bulk theory is $AdS_3$ Chern-Simons
with gauge algebra $\mathfrak {sl}(n) \oplus \mathfrak {sl}(n)$.
This bulk theory  describes gravity coupled to   massless
higher spin fields.
In order to identify the states associated to the physical
graviton it is necessary to specify an embedding of $\mathfrak {sl}(2)$
in $\mathfrak {sl}(n)$. As explained in \cite{Campoleoni:2010zq}, the bulk theory must be
supplemented by suitable boundary conditions in order to guarantee 
an asymptotically $AdS_3$ geometry. The interplay between the
$\mathfrak {sl}(2)$ embedding and these boundary conditions
determines the asymptotic symmetry algebra of the bulk theory,
which is  furnished by two copies (left-moving and right-moving)
of the same
classical infinite-dimensional Poisson algebra.
Interestingly, this physical construction
based on the asymptotic symmetry algebra
is equivalent to 
the classical 
Drinfel'd-Sokolov (DS) Hamiltonian reduction of $\mathfrak {sl}(n)$
associated to the prescribed embedding $\mathfrak{sl}(2) \subset  \mathfrak {sl}(n)$.
In the case of the principal embedding, the outcome of the DS
reduction is the classical $\cW_n$ algebra, whose quantization
yields the quantum  $\cW_n$ algebra.  

If the DS reduction provides the natural way to
get the boundary W-algebra from the bulk Lie algebra,
the notion of {\it wedge algebra}, explored in \cite{Bowcock:1991zk} in great generality,
 proves extremely useful
for proceeding in the opposite direction. 
Let the generators of the W-algebra be denoted as
$W^s(\zeta)$, where $s$ labels the integer holomorphic
dimension of the generator.
Let $W^s_\ell$, $\ell \in\mathbb Z$ be the modes in the Laurent
expansion of $W^s(\zeta)$.
The vacuum preserving modes are  
\beq \label{vacuum_preserving_modes}
W^s_\ell \ , \quad |\ell| < s  \ ,
\eeq
and preserve both the left and right $\mathfrak {sl}(2)$ invariant vacuum.
Our goal is to define a finite-dimensional Lie algebra
generated by the vacuum preserving modes \eqref{vacuum_preserving_modes}.
A na\"ive truncation of the commutators of the original W-algebra
fails in general, due to the non-linear terms that may appear on 
RHS of the commutators of the vacuum preserving modes.
The crucial observation is that, if the W-algebra can be defined
for arbitrary values of the central charge $c$
and satisfies additional non-degeneracy assumptions
listed in  \cite{Bowcock:1991zk}\footnote{The additional assumptions of \cite{Bowcock:1991zk} are that both the quantum W-algebra and its classical limit be positive definite, in particular the metric defined by central terms should be positive definite both in the quantum and classical algebras. These conditions can be presumably relaxed for non-unitary W-algebras --  non-degeneracy of the metric should suffice.}, then all 
non-linear terms on the RHSs of commutators of 
vacuum preserving modes
are suppressed 
in the limit $c \rightarrow \infty$. 
Furthermore, central terms do not contribute
if we restrict to  vacuum preserving modes.
It follows that the algebra becomes linear and, since associativity of the parent W-algebra
holds for any $c$, we are guaranteed to obtain a
{\it bona fide}
 finite-dimensional
Lie algebra satisfying all Jacobi identities.
An essential property of the wedge algebra construction is that,
if the starting point is a W-algebra $\cW^{\rm DS}(\mathfrak{g})$
obtained by DS reduction
of a finite-dimensional Lie algebra $\mathfrak{g}$,
then the wedge algebra of $\cW^{\rm DS}(\mathfrak{g})$
reproduces   $\mathfrak{g}$ itself.
In particular, the wedge algebra of $\cW_n$ is $\mathfrak{sl}(n)$.
Even though we have reviewed a purely bosonic example,
the extension of these considerations to graded Lie algebras
does not pose any essential difficulty.

In our problem, the role of $\cW_n$ is played by the chiral algebra
of $\cN = 4$ SYM with gauge algebra $\su(N)$. 
The existence of this chiral algebra is guaranteed 
if its central charge is tuned to the value 
determined by the cohomological
construction, $c_{\rm 2d} = -3(N^2-1)$.
It is not clear, however,
if for $N\ge 3$ this chiral algebra can be deformed 
to arbitrary $c$.  
As a result, we cannot guarantee the existence of a wedge algebra,
which would be the natural candidate for the sought after
Lie algebra in the bulk.

If we consider the case of infinite $N$, however,
we can infer the existence of a wedge algebra,
which is an ordinary (linear) Lie algebra, albeit infinite dimensional.
The argument relies on large $N$ factorization,
and goes as follows. 
We have established in section 4.1
that, in the large $N$ limit, the supergenerators of the 
chiral algebra are in 1-to-1 correspondence with 
single trace  1/2 BPS operators of $\cN = 4$ SYM theory.
Thanks to the protection ensured by supersymmetry, their
correlators  can be computed in the 
  free field theory limit.
We normalize the fundamental adjoint scalars of $\cN = 4$ SYM
in such a way that their contraction yields schematically
\beq
\langle X^x{}_y \, X^z{}_w \rangle \sim  g_{\rm YM}^2 \, \delta^z_y \, \delta^x_w \ ,
\eeq
where $x,y,z,w$ are fundamental indices of $\mathfrak{su}(N)$,
we   suppressed all spacetime and R-symmetry dependence,
and we restricted to the leading term at large $N$. With the aid of the standard double-line notation, 
it is elementary to  show the following schematic scalings,
\begin{align} \label{scaling_relations}
\langle {\rm tr} X^k \, {\rm tr} X^k \rangle & \sim g_{\rm YM}^{2k} \, N^k = \lambda^k
\ , \nn \\
\langle {\rm tr} X^{k_1} \, {\rm tr} X^{k_2} \, {\rm tr} X^{k_3} \rangle
& \sim g_{\rm YM}^{k_1 + k_2 + k_3} \, N^{-1 + \frac 12(k_1 + k_2 + k_3) }
= \lambda^{\frac 12 (k_1 + k_2 + k_3)} \, N^{-1} \ , \nn \\
\langle : {\rm tr} X^{k_1} \, {\rm tr} X^{k_2} :  \, : {\rm tr} X^{k_1} \, {\rm tr} X^{k_2} : \rangle
& \sim g_{\rm YM}^{2(k_1 + k_2)} \, N^{k_1 + k_2} = \lambda^{k_1 + k_2} \ , \nn \\
\langle {\rm tr} X^{k_1} \, {\rm tr} X^{k_2} \, : {\rm tr} X^{k_3} \, {\rm tr} X^{k_4} : \rangle
& \sim g_{\rm YM}^{k_1 + k_2 + k_3 + k_4} \, N^{\frac 12 (k_1 + k_2 + k_3 + k_4) -2}
= \lambda^{\frac 12 (k_1 + k_2 + k_3 + k_4)} \, N^{-2} \ ,
\end{align}
where $: \; :$ denotes normal ordering and  $\lambda = g_{\rm YM}^2 \, N$ is the 't~Hooft coupling.\footnote{The last scaling relation in \eqref{scaling_relations}
holds for generic $\ell_1$, $\ell_2$, $k_1$, $k_2$.
In the special case $\ell_1 = k_1$, $\ell_2 = k_2$ the  scaling is
$\lambda^{k_1 + k_2} \, N^0$. 
}
If we modify the normalization of the single trace operators, setting
\beq
\cO_k = N \, \lambda^{- \frac 12 k} \,  {\rm tr} X^k  \ ,
\eeq
the previous relations may then be written in the simpler form
\begin{align}
\langle \cO_k \, \cO_k \rangle & \sim N^2
\ , \nn \\
\langle \cO_{k_1} \, \cO_{k_2}  \, \cO_{k_3} \rangle
& \sim N^2 \ , \nn \\
\langle : \cO_{k_1} \, \cO_{k_2} :  \, : \cO_{k_1} \, \cO_{k_2} : \rangle
& \sim N^4  \ , \nn \\
\langle \cO_{\ell_1} \, \cO_{\ell_2} \, : \cO_{k_1} \, \cO_{k_2} : \rangle
& \sim N^2 \ .
\end{align}
These relations constrain the $N$ dependence of the OPE coefficients
in the OPE of two $\cO_k$ operators. 
Very schematically, we   may then write
\beq
\cO_{k_1} \, \cO_{k_2} \sim N^2 \, \delta_{k_1 k_2}   \, \mathbb I + N^0 \,
C^{k_3}_{k_1k_2} \cO_{k_3}  + N^{-2} \,  C^{k_3k_4}_{k_1k_2}  : \cO_{k_3} \, \cO_{k_4} : + \dots
\eeq
where we have only kept track of the $N$ dependence.
Furthermore we have only focused on potential {\it singular} terms in the OPE,
and in particular we supposed $(k_3,k_4) \neq(k_1,k_2)$.\footnote{In
 the case $(k_3,k_4) =(k_1,k_2)$ the $N$ scaling is $N^0$
due to the remark in the previous footnote. This is in accordance with the
tautological observation that $:\cO_{k_1} \, \cO_{k_2}:$ enters
the {\it regular} part of the $\cO_{k_1} \, \cO_{k_2} $ OPE with coefficient one.
} 
As we can see, if double trace operators enter the singular part
of the OPE of two single trace operators, the corresponding OPE coefficient
is suppressed by a power of $N^{-2}$. It is not hard to convince oneself that this
pattern persists for all multi-trace operators: 
if a trace-$m$ operator enters the singular part of the OPE,
it appears with a power $N^{2-2m}$.
This argument implies that all non-linear terms
in the chiral algebra must be suppressed at large $N$.
As a result, the obstruction to the consistency of the 
wedge algebra generated by the vacuum preserving modes
is removed, and we obtain a well-defined, infinite-dimensional
Lie algebra.
This is our candidate for the higher-spin Lie algebra in the bulk.

All the necessary information  for determining the structure
constants of this Lie algebra is  contained    in the OPE
of single-trace operators in the $1/2$ BPS chiral ring of  $\cN = 4$ SYM.
It would be desirable, however, to have a more direct construction
of this higher-spin algebra, along the lines of \cite{Gaberdiel:2012uj} in the context
of minimal model holography.
Such investigation is left for future work. Let us list here the expected vacuum preserving modes 
of the operators in \eqref{psu112_pattern} that generate the wedge Lie algebra,
suppressing   $\mathfrak {su}(2)_F$ indices
for simplicity:
\beq
\begin{array}{clllll}
\phantom{mmmmmmmmm}
&&\phantom{mmmm}  \text{$n$ even:}
\phantom{mmmmmmmmm} &
& \phantom{mmmm} \text{$n$ odd:} \\[2mm]
\mathsf J^{(n)}_\ell &  \ell &=0, \pm 1, \dots  , \pm \tfrac 12 n \ , &
 \ell &=\pm \tfrac 12,
\pm \tfrac 32 \ ,  \dots  , \pm \tfrac 12 n  \ ,\\[1mm]
\mathsf G^{(n)}_\ell   , \tilde{ \mathsf G}^{(n)}_\ell  & 
\ell &= \pm \tfrac 12, 
\pm \tfrac 32, \dots  , \pm (\tfrac 12 +\tfrac 12 n)  \ , & 
\ell &= 0, 
\pm 1, \dots  , \pm (\tfrac 12 +\tfrac 12 n)   \ , \\[1mm]
\mathsf T^{(n)}_\ell   & 
\ell &=0, \pm 1, \dots \pm (1+\tfrac 12 n)  , \ & 
\ell &=\pm \tfrac 12, 
\pm \tfrac 32, \dots \pm (1+\tfrac 12 n)  \ .
\end{array} \nn
\eeq

\section{Discussion}
In this work we have addressed the problem of determining the
holographic dual of the protected chiral algebra
of $\cN = 4$ SYM theory with gauge algebra $\mathfrak {su}(N)$
in the large $N$ limit.
The resulting picture is the following. The cohomological construction
on the field theory side is mirrored by supersymmetric 
localization in the bulk. By virtue of this localization, type IIB supergravity
on $AdS_5 \times S^5$ reduces to a Chern-Simons theory
defined on an $AdS_3$ slice of the $AdS_5$ space.
The gauge algebra of the Chern Simons theory is an infinite-dimensional
supersymmetric
higher spin Lie algebra, whose structure can {\it a priori} be extracted
from the  coefficients in the OPE of the single-trace $1/2$ BPS generators
of the chiral ring of $\cN = 4$ SYM theory.

Although we were not able to provide a proof for all 
aspects of the above picture, we have collected several pieces of evidence
in favor of it. To begin with, we have implemented the localization program
explicitly in a simplified setup, illustrating how an $AdS_3$ Chern-Simons theory
emerges non-trivially from a five-dimensional gauge theory on $AdS_5$.
Secondly, we have established the super chiral algebra conjecture of \cite{Beem:2013sza}
in the case of infinite $N$, providing the correspondence between   
supergenerators of the chiral algebra, 
single-trace $1/2$ BPS generators of the chiral ring of $\cN  =4$ SYM,
and Kaluza-Klein modes of type IIB supergravity on $AdS_5 \times S^5$.
Finally, we have identified a natural candidate for the higher-spin algebra
in which the Chern-Simons connection takes values.
It is the wedge algebra of the chiral algebra,
{\it i.e.~}the (infinite-dimensional) Lie algebra generated by the vacuum
preserving modes of the generators of the chiral algebra.
We furnished an argument based on large $N$ factorization
for the existence of this wedge algebra, and we have connected its
structure to the OPEs of single-trace $1/2$ BPS scalar operators of $\cN = 4$ SYM theory.

It is interesting to contrast this four-dimensional setup to the
six-dimensional case in which the superconformal field theory is the $(2,0)$ theory
of type $A_{N-1}$.
As established in \cite{Beem:2014kka}, the protected chiral algebra in this case
coincides with $\cW_{N}$.
The latter is defined for arbitrary values of the central charge $c$
and admits $\mathfrak {sl}(N)$ as its wedge algebra.
The gravity dual of a $\cW_{N}$ chiral algebra is 
thus an $AdS_3$ Chern-Simons theory with gauge algebra   $\mathfrak {sl}(N)$.
These facts are well-known in the context of minimal model
holography \cite{Gaberdiel:2012uj}, and the large $N$ limit is also 
well understood. 
This problem is thus simpler from a bottom-up point of view.
From a top-down perspective, however, this case is considerably
more complicated. In the large $N$ limit we can access the holographic
dual
of the $(2,0)$ theory of type $A_{N-1}$ via eleven-dimensional
supergravity on $AdS_7 \times S^4$. In contrast to the case
studied in this paper, it is not possible to single out a simplified setup
without dynamical gravity to perform the localization computation.
As a result, a direct check of the emergence of the claimed Chern-Simons
theory would require a full-fledged localization computation in supergravity.

The cohomological construction of the protected chiral algebra in $4d$ SCFTs
also has a counterpart for $3d$ SCFTs \cite{Chester:2014mea, Beem:2016cbd}: the protected sector
gives rise to a one-dimensional topological algebra.
The construction requires at least $\cN = 4$ in three dimensions,
and is in particular applicable to the maximally supersymmetric case $\cN = 8$.
In the latter situation the holographic dual can be accessed via 
eleven-dimensional supergravity on $AdS_4 \times S^7$. In analogy to the
case discussed in this work, it is possible to single out a simplified model,
involving the dynamics of a vector multiplet in $AdS_4$, in order to perform
the localization computation. 
The outcome of the localization procedure is expected to live on an $AdS_2$ slice
of $AdS_4$, and it would be interesting to show this explicitly.

A general formalism for 
defining   twisted supergravity theories
 has been recently introduced in~\cite{Costello:2016mgj}, with the motivation to discuss twisted versions of the AdS/CFT correspondence. It would be extremely interesting to apply this
 formalism to our setup.

\acknowledgments
Our work is  supported in part by NSF Grant PHY-1316617.
We are grateful to Chris Beem, Cyril Closset, Stefano Cremonesi, Carlo Meneghelli, Wolfger Peelaers, and Balt van Rees for useful conversations.


\appendix


\section{Conventions and technical material}
\label{app:conventions}

\subsection{Gamma matrices in various dimensions}
\label{app:gamma_matrices}

Gamma matrices in Euclidean $5d$ dimensions are hermitian and satisfy
\beq
\{ \gamma^{\hat \mu} , \gamma^{\hat \nu} \} = 2 \delta^{\hat \mu \hat \nu} \, \mathbb I_4 \ , 
\qquad
C_5 \, \gamma^{\hat \mu} \, C_5^{-1} = (\gamma^{\hat \mu})^{\sf T} \ , \qquad
C_5^{\sf T} = - C_5 \ , \qquad
\gamma^{1}\gamma^2\gamma^3\gamma^4\gamma^5 =\mathbb I_4 \ ,
\eeq
where $\hat \mu, \hat \nu = 1, \dots 5$ are flat spacetime indices and $C_5$
is the unitary charge conjugation matrix.
We also need gamma matrices for the R-symmetry directions
$\cI = 6,\dots,9,0$ with signature $(4,1)$. They satisfy
\beq \label{Rsymmetry_gammas}
\{ \rho^\cI , \rho^\cJ \} = 2 \eta^{\cI \cJ} \, \mathbb I_4  \ , 
\qquad
C_{4,1} \, \rho^\cI \, C_{4,1}^{-1} = (\rho^\cI)^{\sf T} \ , \qquad
C_{4,1}^{\sf T} = - C_{4,1} \ , \qquad
\rho^6 \rho^7 \rho^8 \rho^9 \rho^0 =i\, \mathbb I_4 \ ,
\eeq
where $\eta_{IJ} = {\rm diag}(+^4,-)$, $\rho^0$ antihermitian, and the other
$\rho^\cI$ hermitian, $C_{4,1}$ unitary.

Let us combine these objects to obtain a convenient representation
of the chiral 
  $16\times 16$ blocks $\Gamma^M$, $\tilde \Gamma^M$ of gamma matrices in ten
dimensions, where $M = 1, \dots, 9,0$ and the flat metric
is $\eta_{MN} = {\rm diag}(+^9,-)$.
Let us set
\begin{align} \label{10d_gammas}
\Gamma^{\hat \mu} &= C_5 \gamma^{\hat \mu} \otimes C_{4,1} \ , 
&
\Gamma^\cI &= -i\, C_5 \otimes C_{4,1} \rho^\cI \ , \nn \\
 \tilde \Gamma^{\hat \mu} &= \gamma^{\hat \mu} C_5^{-1} \otimes C_{4,1}^{-1} \ ,
&
\tilde \Gamma^\cI &= i\, C_5^{-1} \otimes \rho^\cI C_{4,1}^{-1} \ .
\end{align}
We may then check the Clifford algebra relations and symmetry properties
\beq
\Gamma^{(M} \tilde \Gamma^{N)} = \eta^{MN} \, \mathbb I_{16} \ , \qquad
\tilde \Gamma^{(M}   \Gamma^{N)} = \eta^{MN} \, \mathbb I_{16} \ , \qquad
(\Gamma^M)^{\sf T} = \Gamma^M \ ,\qquad
(\tilde \Gamma)^{\sf T} = \tilde \Gamma^M \ ,
\eeq
as well as the chirality relations
\beq
\tilde \Gamma^1   \Gamma^2 
\tilde \Gamma^3   \Gamma^4
\tilde \Gamma^5   \Gamma^6 
\tilde \Gamma^7   \Gamma^8 
\tilde \Gamma^9   \Gamma^0  =  \mathbb I_{16} \ ,
 \qquad 
\Gamma^1 \tilde \Gamma^2 
\Gamma^3 \tilde \Gamma^4
\Gamma^5 \tilde \Gamma^6 
\Gamma^7 \tilde \Gamma^8 
\Gamma^9 \tilde \Gamma^0  = - \mathbb I_{16} \ .
\eeq
These relations imply that $\Gamma^M$ maps positive-chiarality
spinors to negative-chirality spinors, while $\tilde \Gamma^M$
acts in the opposite way.
Let us assign a lower Weyl index $\alpha = 1, \dots, 16$ to a positive-chirality
spinor $\Psi_\alpha$, and an upper index to a spinor of negative chirality
$\tilde \Psi^\alpha$.
It follows that the index structure of gamma matrices is 
 $\tilde \Gamma^M_{\alpha \beta}$,
$\Gamma^{M \alpha \beta}$. 
Using Weyl indices we can conveniently
formulate the ``triality identities"   as
\beq
 \Gamma^{M (\alpha \beta} \, \Gamma_M{}^{\gamma) \delta} = 0 \ , \qquad
\tilde \Gamma^{M}{}_{ (\alpha \beta} \, \tilde \Gamma_M{}_{\gamma) \delta} = 0 \ .
\eeq
Let us stress that the Lorentz generators are one half of the matrices
\beq
\Gamma^{MN} = \tilde \Gamma^{[M} \Gamma^{N]}  \ , \qquad 
\tilde \Gamma^{MN} =   \Gamma^{[M} \tilde  \Gamma^{N]} \ ,
\eeq
and since  $(\tilde \Gamma^{MN})^{\sf T} = -\Gamma^{MN}$
 positive and negative chirality representations are dual.
As a result, Majorana bilinears are simply built contracting Weyl indices
on $\Psi_\alpha$, $\tilde \Psi^\alpha$, 
 $\tilde \Gamma^M_{\alpha \beta}$,
$\Gamma^{M \alpha \beta}$.

As a final remark, 
recall that we adopt an off-shell supersymmetry formalism
that
realizes manifestly only a subalgebra $\mathfrak{so}(2)_R \times \mathfrak{so}(2,1)_R$
of the full R-symmetry group $\mathfrak{so}(4,1)_R$,
associated to the split $\cI = (i,A)$, $i = 6,7$, $A = 8,9,0$.
It is therefore convenient to specify further 
the representation of gamma matrices $\rho^\cI$ by 
requiring
\beq \label{rho890}
\rho^8 \rho^9 \rho^0 = \begin{pmatrix}
\mathbb I_2 & 0 \\
0 & - \mathbb I_2
\end{pmatrix} \ .
\eeq
An explicit realization of all the gamma matrices 
considered in this section in terms of Pauli matrices
can be found in subsection \ref{app:explicit_gammas}
below.

\subsection{Embedding formalism for $AdS_5$ and Killing spinors}
\label{app:Killing_spinors}

Let us review some well-known constructions in order to fix
our notations and conventions.
We realize Euclidean $AdS_5$ as a hyperboloid
in the embedding space $\mathbb R^{5,1}$.
Let coordinates on the latter be $X^\cA$, 
in which the $6d$ vector index $\cA$ 
is split as $\cA \rightarrow (a, \underline 0, \underline 5)$,
with $a = 1, \dots, 4$.
The $a$ indices 
correspond to  directions along the
conformal boundary of $AdS_5$, while $\underline 0$, $\underline 5$
 are an auxiliary timelike
and spacelike directions respectively
(we use underlined indices
to avoid possible confusion with other
spacetime or R-symmetry indices).
The hyperboloid equation is
\beq
\eta_{\cA \cB} X^\cA X^\cB =
 \delta_{ab} X^a X^b - X^{\underline 0}X^{\underline 0}
 + X^{\underline 5}X^{\underline 5}  = - R^2 \ ,
\eeq
where $R$ is the radius of $AdS_5$. The relation between
the embedding coordinates $X^\cA$ and the Poincar\'e coordinates
$x^a$, $z$ in \eqref{AdSmetric} is
\beq
X^a = \frac{R x^a}{z} \ , \qquad
X^{\underline 0} + X^{\underline 5} = \frac{x^a x_a + z^2}{z} \ , \qquad
X^{\underline 0} - X^{\underline 5} = \frac{R^2}{z} \ .
\eeq

In the discussion of $AdS_5$ Killing spinors it is convenient to adopt
a representation of $5d$ gamma matrices $\gamma^{\hat \mu}$
derived from a suitable chiral representation of gamma matrices in
the $\mathbb R^{5,1}$ embedding space.
Let $\tau^\cA$, $\tilde \tau^\cA$ be chiral blocks of gamma matrices
of $\mathbb R^{5,1}$ satisfying
\begin{align}
& \tau^{(\cA} \tilde \tau^{\cB)} = \eta^{\cA
\cB} \, \mathbb I_4 \ , \qquad
\tilde \tau^{(\cA}   \tau^{\cB)} = \eta^{\cA
\cB} \, \mathbb I_4 \ , \qquad
(\tau^{\cA})^{\sf T} = - \tau^{\cA} \ , \qquad
(\tilde \tau^{\cA})^{\sf T} = -\tilde \tau^{\cA}   \\[2mm]
&\tilde \tau^1 \tau^2 \tilde \tau^3 \tau^4 \tilde \tau^{\underline 0} \tau^{\underline 5}
=  \mathbb I_4 \ , \qquad
  \tau^1 \tilde \tau^2   \tau^3 \tilde \tau^4   \tau^{\underline 0}
    \tilde \tau^{\underline 5}
= - \mathbb I_4  \ , 
\qquad
(\tau^a)^\dagger = \tilde \tau^a \ , \qquad
(\tau^{\underline 5})^\dagger = \tilde \tau^{\underline 5} \ , \qquad
(\tau^{\underline 0})^\dagger = - \tilde \tau^{\underline 0} \ . \nn
\end{align}
Let us construct the $5d$ gamma matrices 
$\gamma^{\hat \mu}=(\gamma^a , \gamma^5)$
and $C_5$ from
these objects via
\beq \label{building_5dgammas}
\gamma^{a} = \tilde \tau^{[a}   \tau^{\underline 0]} \ , \qquad
\gamma^5 = \gamma^1 \gamma^2 \gamma^3 \gamma^4 \ , \qquad
C_5 =  i \,   \tau^{\underline 0} \ .
\eeq
The first relation identifies the vector space
of $5d$ Dirac 4-component spinors
with 4-component 
Weyl spinors of positive chirality 
in the $\mathbb R^{5,1}$  embedding space. 
The matrix $\gamma^5$ is   associated
to the radial coordinate $z$, but at the same time plays 
the role of chirality matrix along the
conformal boundary of $AdS_5$.

Let $\underline \lambda$, $\tilde{ \underline \lambda}$ be  constant, 
4-component 
Weyl spinors in the embedding space $\mathbb R^{5,1}$
of positive and negative chirality, respectively.
Their $SO(5,1)$ transformation laws read
\beq \label{lambda_rotations}
\delta \underline \lambda = 
\ell \cdot \underline \lambda \equiv
\tfrac 14 \ell_{\cA \cB}  \,
\tilde \tau^{[\cA}
\tau^{\cB]} \, \underline \lambda\ , \qquad
\delta \tilde{ \underline \lambda} =
\ell \cdot \tilde{\underline \lambda }
\equiv \tfrac 14 \ell_{\cA \cB}  \,
  \tau^{[\cA}
\tilde \tau^{\cB]} \, \tilde { \underline \lambda}  \ ,
\eeq
where $\ell_{[\cA \cB]}$ is the infinitesimal parameter.
Let us consider the maps
\begin{align} 
\underline \lambda \;\; \mapsto \;\;
\psi(\underline \lambda) &=
\frac{R}{\sqrt z} \left[
 \frac{\mathbb I_4 - \gamma^5}{2}
-\frac{x^a \gamma_a + z \, \mathbb I_4}{R}  \;   \frac{\mathbb I_4 + \gamma^5}{2}
\right] \underline \lambda \ ,    \label{Killing_spinor_map1}
\\
\tilde {\underline \lambda} \;\; \mapsto \;\;
\tilde \psi(\tilde{\underline \lambda}) &=
\frac{R}{\sqrt z} \left[
 \frac{\mathbb I_4 + \gamma^5}{2}
+ \frac{x^a \gamma_a - z \, \mathbb I_4}{R} \;   \frac{\mathbb I_4 - \gamma^5}{2}
\right] \, \tilde \tau^{\underline 0} \, 
\tilde {\underline \lambda }\ . \label{Killing_spinor_map2}
\end{align}
First of all, $\psi(\underline \lambda)$, $\tilde \psi(\tilde{\underline \lambda})$
satisfy the $5d$ Killing spinor equations
\beq  \label{5d_KSE}
\nabla_\mu \psi = \frac {1}{2R} \gamma_\mu \psi \ , \qquad
\nabla_\mu \tilde \psi =- \frac {1}{2R} \gamma_\mu \tilde \psi \ ,
\eeq
provided we adopt the vielbein \eqref{vielbein} with the radial coordinate
$z$ associated to $\gamma^5$.
Second of all, the maps \eqref{Killing_spinor_map1},
\eqref{Killing_spinor_map2}
are equivariant under the action of $\mathfrak {so}(5,1)$.
At the infinitesimal level we can write
\beq
\psi (\underline \lambda + \ell \cdot \underline \lambda)
= \psi(\underline \lambda) + \ell \cdot \psi(\underline \lambda) \ ,
\eeq
where the action $\ell \cdot \underline \lambda$
is defined in \eqref{lambda_rotations},
while the action $ \ell \cdot \psi(\underline \lambda)$
is realized as the negative of the   
Lie derivative along the vector field $\xi^\mu$
associated to   $\ell$,
\beq
 \ell \cdot \psi(\underline \lambda)
\equiv - \cL_{\xi } \psi(\underline \lambda)
= - \xi^\mu \nabla_\mu \psi(\underline \lambda)
- \frac 14 \nabla_\mu \xi_\nu \, \gamma^{\mu\nu}  \psi(\underline \lambda) \ ,
\qquad
\delta X^{\cA} = \ell^{\cA}{}_{\cB} X^{\cB} \ , \quad
\delta x^\mu = \xi^\mu \ .
\eeq
Completely analogous statements hold for the map
$\tilde {\underline \lambda} \mapsto \tilde \psi(
\tilde {\underline \lambda}
)$.

In order to implement the 
 off-shell supersymmetry formulation of \cite{Minahan:2015jta}
followed in the main text,
we have to promote the 4-component $AdS_5$
Killing spinors $\psi(\underline \lambda)$,
$\tilde \psi(\tilde {\underline \lambda})$ to 
suitable solutions to 
 \eqref{KSE}, which holds for 16-component
spinors in ten dimensions and therefore  implicitly  involves the
R-symmetry directions.
Following our realization of $10d$ gamma matrices \eqref{10d_gammas}
we seek solutions to \eqref{KSE} of the form
\beq
\epsilon(\underline \lambda, \eta) = \psi(\underline \lambda) \otimes \eta \ , \qquad
\tilde \epsilon(\tilde {\underline \lambda}, \tilde \eta)
 =\tilde \psi( \tilde {\underline \lambda}) \otimes \tilde \eta \ ,
\eeq
where $\eta$, $\tilde \eta$ are constant 4-component
spinors associated to the gamma matrices $\rho^\cI$ of the
R-symmetry space. Using the definition of $\Lambda$ in \eqref{def_Lambda}
and the $5d$ Killing spinor equations \eqref{5d_KSE}
we see that $\epsilon(\underline \lambda, \eta)$,
$\tilde \epsilon(\tilde {\underline \lambda}, \tilde \eta)$
solve \eqref{KSE} as soon as
the constant spinors $\eta$, $ \tilde  \eta$
satisfy the   constraints
\beq \label{eta_constraints}
\rho^8 \rho^9 \rho^0 \, \eta = \eta \ , \qquad
\rho^8 \rho^9 \rho^0 \, \tilde \eta = -  \tilde \eta \ .
\eeq
It follows that $\epsilon(\underline \lambda, \eta)$,
$\tilde \epsilon(\tilde {\underline \lambda}, \tilde \eta)$
obey the $10d$   constraints
\beq
\tilde \Gamma^6 \Gamma^7 \, \epsilon(\underline \lambda, \eta)
= i \, \epsilon(\underline \lambda, \eta) \ , \qquad
\tilde \Gamma^6 \Gamma^7 \, \tilde \epsilon(\tilde {\underline \lambda}, \tilde \eta)  = - i \, \tilde \epsilon(\tilde {\underline \lambda}, \tilde \eta) \ ,
\eeq
readily checked using the last equation in \eqref{Rsymmetry_gammas}.
The space of solutions to \eqref{KSE} is thus 
explicitly
decomposed into eigenspaces of the $\mathfrak{so}(2)_R$ 
hermitian generator
$\tfrac i2 \tilde \Gamma^6 \Gamma^7$.
This decomposition mirrors the 
partition of the boundary supercharges into
$\cQ^{xI} \equiv (\cQ^I_\alpha , \tilde \cS^{I \dot \alpha})$
and $\tilde \cQ_{xI} \equiv ( \cS_I^\alpha, \tilde \cQ_{I\dot\alpha} )$,
where here $\alpha = +,-$, $\dot \alpha = \dot+, \dot-$ are Weyl indices in four dimensions,
which can be collected into an index $x = 1, \dots, 4$
in the (anti)fundamental of the
conformal algebra  $\mathfrak{su}^*(4) \cong \mathfrak{so}(5,1)$.

Let us consider the canonical basis $\{ \underline \lambda_{(x)} \}$,
$x = 1,\dots,4$ of positive-chirality Weyl spinors in the embedding
space $\mathbb R^{5,1}$, and its counterpart 
$\{ \tilde {\underline \lambda}{}^{(x)} \}$
for negative chirality,
given by
\beq
\underline \lambda_{(1)} = \tilde {\underline \lambda}{}^{(1)} 
= \begin{pmatrix}
1 \\ 0 \\ 0 \\ 0
\end{pmatrix} \ , \dots \ , \;
\underline \lambda_{(4)} =\tilde {\underline \lambda}{}^{(4)} 
= \begin{pmatrix}
0 \\ 0 \\ 0 \\ 1
\end{pmatrix} \ .
\eeq
In light of \eqref{rho890} we also consider the objects
 $\eta_{(I)}$, $\tilde \eta{}^{(I)}$,
$I = 1,2$  
\beq
\eta_{(1)} 
= \begin{pmatrix}
1 \\ 0 \\ 0 \\ 0
\end{pmatrix} \ , \qquad
\eta_{(2)} 
= \begin{pmatrix}
0 \\ 1 \\ 0 \\ 0
\end{pmatrix} \ , \qquad
\tilde \eta^{(1)} 
= \begin{pmatrix}
0 \\ 0 \\ 1 \\ 0
\end{pmatrix} \ , \qquad
\tilde \eta^{(2)} 
= \begin{pmatrix}
0 \\ 0 \\ 0 \\ 1
\end{pmatrix} \ .
\eeq
We may then set
\beq
\epsilon_{(xI)} = \epsilon(\underline \lambda_{(x)} , \eta_{(I)}) \ , \qquad
\tilde \epsilon^{(xI)} = \tilde \epsilon(
\tilde{\underline \lambda}{}^{(x)} , \tilde \eta^{(I)}) \ ,
\eeq
and thus obtain a canonical basis of solutions to \eqref{KSE}.
Thanks to the equivariance of the maps \eqref{Killing_spinor_map1},
\eqref{Killing_spinor_map2}, the label $x$ can be regarded
as a bona fide (anti)fundamental index of 
$\mathfrak {su}^*(4)$.
By the same token, $I$ can be regarded as an index 
in the fundamental of $\mathfrak{sl}(2,\mathbb R)_R \cong \mathfrak{so}(2,1)_R$.
The most general Killing spinor in ten dimensions
may then be written as
\beq
\epsilon = \kappa^{(xI)} \, \epsilon_{(xI)}  + \tilde \kappa_{(xI)} \, 
\tilde \epsilon^{(xI)} \ ,
\eeq
where the constants $\kappa$, $\tilde \kappa$
are in one-to-one correspondence
with the boundary supercharges
$\cQ^{xI}$, $\tilde \cQ_{xI}$.
In the implementation of the localization technique
in section \ref{sec:localization}
we select the spinor specified by
\beq
\kappa^{(21)} = \tfrac i2   \ , \qquad
\kappa^{(42)} = \tfrac 12   \ , \qquad
\tilde \kappa_{(22)} = - \tfrac i2  \ , \qquad
\tilde \kappa_{(41)} =- \tfrac 12   \ , 
\eeq	
with all other constants $\kappa$, $\tilde \kappa$
vanishing. 

\subsection{Some explicit expressions}
\label{app:explicit_gammas}

We adopt the following representation for the chiral 
blocks $\tau^\cA$, $\tilde \tau^\cA$ of gamma matrices
in the embedding space $\mathbb R^{1,5}$,
\beq
\begin{array}{llrcc}
\tau^1 & = & -i \,  \sigma_2 & \otimes& \sigma_1 \ ,  \\
\tau^2 & = & -    \sigma_1& \otimes& \sigma_2 \ , \\
\tau^3 & = & i \,  \sigma_2 &\otimes &\sigma_3 \ , \\
\tau^4 & = &     \sigma_2& \otimes& \mathbb I_2 \ , \\
\tau^{\underline 0} & = &   \mathbb I_2 & \otimes &\sigma_2 \ , \\
\tau^{\underline 5} & = &    \sigma_3 &\otimes &\sigma_2 \ , 
\end{array} 
\hspace{2cm}
\begin{array}{llrcc}
\tilde \tau^1 & = & i \, \sigma_2 & \otimes& \sigma_1 \ , \\
\tilde \tau^2 & = & -    \sigma_1& \otimes& \sigma_2 \ , \\
\tilde \tau^3 & = & -i \, \sigma_2 &\otimes &\sigma_3 \ , \\
\tilde \tau^4 & = &     \sigma_2& \otimes& \mathbb I_2 \ , \\
\tilde \tau^{\underline 0} & = & -   \mathbb I_2 & \otimes &\sigma_2 \ , \\
\tilde \tau^{\underline 5} & = &    \sigma_3 &\otimes &\sigma_2 \ .
\end{array}
\eeq
This representation determines the form of spacetime $5d$ gamma matrices
according to \eqref{building_5dgammas},
\beq
\begin{array}{llrcc}
\gamma^1 & = & -    \sigma_2 & \otimes& \sigma_3 \ ,  \\
\gamma^2 & = & -    \sigma_1& \otimes& \mathbb I _2 \ , \\
\gamma^3 & = & -   \sigma_2 &\otimes &\sigma_1 \ , \\
\gamma^4 & = &     \sigma_2& \otimes& \sigma_2 \ , \\
\gamma^5 & = & -  \sigma_3 & \otimes &\mathbb I _2 \ , 
\end{array}  
\hspace{2cm}
C_5 = \mathbb I_2 \otimes i\, \sigma_2 \ .
\eeq
We use a different representation for the gamma matrices
$\rho^\cI$
 in the 
R-symmetry directions,
which is better suited for the split $\cI = (i,A)$,
$i = 6,7$, $A = 8,9,0$, and satisfies \eqref{rho890},
\beq
\begin{array}{llrcc}
\rho^6 & = &      \sigma_1 & \otimes& \mathbb I _2 \ ,  \\
\rho^7  & = &      \sigma_2 & \otimes& \mathbb I _2 \ , \\
\rho^8  & = &     \sigma_3  &\otimes &\sigma_1 \ , \\
\rho^9  & = &     \sigma_3 & \otimes& \sigma_2 \ , \\
\rho^0  & = &  -i \,  \sigma_3 & \otimes &\sigma_3 \ , 
\end{array} 
\hspace{2cm}
C_{4,1} = i\, \sigma_1 \otimes \sigma_2 \ .
\eeq
The explicit expression for the 16-component Killing spinor
$\epsilon$ is
\beq \label{our_explicit_epsilon}
\epsilon =\frac{1}{2 \sqrt z} ( 0 \; , \; 
- i \rho e^{i \varphi} \; , \;
0 \; , \;
-z \; , \;
i R \; , \;
i \bar \zeta \; , \;
0 \; , \;
0 \; , \;
0 \; , \;
0 \; , \;
- i R \; , \;
- i \bar \zeta \; , \;
0 \;  , \;
-z \;  , \;
0 \;  , \;
- i \rho^{- i   \varphi}
)^{\sf T} \ .
\eeq

\subsection{More on the auxiliary pure spinors}
\label{app:auxiliary}

The supersymmetry transformations are presented using an orthonormal
set of pure spinors satisfying
\beq \label{both_nu_equations}
\nu_m \Gamma^M \epsilon = 0 \ , \qquad
\nu_m \Gamma^M \nu_n = \delta_{mn} \, \epsilon \Gamma^M \epsilon  \ .
\eeq
For our choice of Killing spinor $\epsilon$ there is no simple closed form
for such a set of orthonormal pure spinors $\nu_m$. We can, however, relax the orthonormality condition
and impose the first relation only.
The index $  m$ can now be regarded as a $\mathfrak{gl}(7)$ index.
For any such set of spinors we have
\beq \label{cM_equation}
\nu_{  m} \Gamma^M \nu_{  n} = \cM_{  m   n} \,
 \epsilon \Gamma^M \epsilon \ ,
\eeq
where the symmetric matrix $\cM_{  m   n}$ is non-degenerate
and is used to raise and lower $  m$ indices.

We can achieve formal $\mathfrak{gl}(7)$ covariance by introducing a 
background $\mathfrak{gl}(7)$ connection $Q_\mu {}^{  m} {}_{  n}$
and utilizing  covariant derivatives such as
\begin{gather}
\mathscr D_\mu \nu_{  m} = \nabla_\mu  \nu_{  m}
- Q_\mu {}^{  n} {}_{  m} \, \nu_{  n} \ , \quad
\mathscr D_\mu K^{  m} = D_\mu K^{  m}
+ Q_\mu {}^{  m} {}_{  n} \, K^{  n} \ , \nn \\ 
\mathscr D_\mu \cM_{mn} = \partial_\mu \cM_{mn}
- Q_\mu {}^p{}_m \, \cM_{pn}
- Q_\mu {}^p{}_n \, \cM_{mp} \ .
\end{gather}
The $\mathfrak{gl}(7)$ transformation laws of $\nu$, $K$, $\cM$, $Q$ are 
\begin{align}
& K'{}^{  m} =  M^{  m} {}_{  n} \, K^{  n}  \ , \qquad
\nu'_{  m} = \nu_{  n} \, (M^{-1})^{  n}{}_{  m} \ , 
\qquad
\cM'_{mn} = \cM_{pq} \, (M^{-1})^{  p}{}_{  m}(M^{-1})^{  q}{}_{  n} \ ,
 \nn \\
&Q'_\mu {}^{  m}{}_{  n}
= M^{  m} {}_{  r} \, Q_\mu {}^{  r}{}_{  s} \,
(M^{-1})^{  s} {}_{  n} 
+ M^{  m} {}_{  r} \, \partial_\mu
(M^{-1})^{  s} {}_{  n}  \ ,\qquad M \in \mathfrak{gl}(7) \ .
\end{align}
All equations used in main text 
can be   $\mathfrak{gl}(7)$-covariantized
raising and lowering $m$ indices with $\cM_{mn}$ rather than $\delta_{mn}$
and
replacing $\nabla_\mu$, $D_\mu$ with $\mathscr D_\mu$ given above
whenever acting on $\nu_m$, $K^m$.
Equivalently, we can regard the equations in the main text as
written in a specific class of $\mathfrak{gl}(7)$ frames for which
\beq
\cM_{  m   n} = \delta_{mn} \ , \qquad
Q_\mu {}^m {}_n = 0 \ .
\eeq
It is clear that, in such a frame,
\beq \label{cov_const_cM}
\mathscr D_\mu \cM_{mn} = 0 \ ,
\eeq
but since this equation is covariant it holds in any $\mathfrak{gl}(7)$ frame.
From a practical point of view, 
one first finds a basis of 
unnormalized pure spinors satisfying 
the first equation in \eqref{both_nu_equations} and then
computes the associated $\cM_{mn}$ via \eqref{cM_equation}.
One may then compute $Q_\mu {}^m {}_n$ from \eqref{cov_const_cM}
upon imposing the gauge-fixing condition
\beq
\cM_{[m| p} \, Q_\mu {}^p {}_{n]} = 0  \ ,
\eeq
although we do not need the explicit expression for $Q_\mu {}^p{}_{n}$
in the implementation of localization.

Let us now record a viable set of $\nu$ spinors for
our Killing spinor $\epsilon$ given in \eqref{our_explicit_epsilon}. 
We have
\beq
\begin{array}{cccccccccccccccccccc}
\nu_1 & = \frac{1}{\sqrt R}& 
( &
0 ,& 
\rho \, e^{i \varphi} ,& 
0 ,& 
0 ,& 
R ,& 
0 ,& 
0 ,& 
0 ,& 
0 ,& 
0 ,& 
-R ,& 
0 ,& 
0 ,& 
0 ,& 
0 ,& 
\rho \, e^{- i\varphi} &
) ^{\sf T} ,\\[2mm]
\nu_2 & = \frac{1}{\sqrt R} & 
( &
0 ,& 
-i \, z \, e^{2i \varphi} , & 
0 ,& 
0 ,& 
0 ,& 
0 ,& 
R ,& 
0 ,& 
- R \, e^{2i\varphi} , & 
0 ,& 
0 ,& 
0 ,& 
0 ,& 
0 ,& 
0 ,& 
-i\,z  &
) ^{\sf T} ,\\[2mm]
\nu_3 & = \frac{1}{\sqrt R} & 
( &
R ,& 
\bar \zeta ,& 
0 ,& 
0 ,& 
0 ,& 
0 ,& 
0 ,& 
0 ,& 
0 ,& 
0 ,& 
0 ,& 
0 ,& 
0 ,& 
0 ,& 
R\, e^{-2i\varphi} , & 
\bar \zeta \, e^{-2i\varphi} &
) ^{\sf T} ,\\[2mm]
\nu_4 & = \frac{1}{\sqrt R} & 
( &
0 ,& 
0 ,& 
0 ,& 
R ,& 
0 ,& 
0 ,& 
0 ,& 
0 ,& 
0 ,& 
0 ,& 
0 ,& 
0 ,& 
0 ,& 
R ,& 
0 ,& 
0 &
) ^{\sf T} ,\\[2mm]
\nu_5 & = \frac{1}{\sqrt R} & 
( &
0 ,& 
0 ,& 
0 ,& 
0 ,& 
0 ,& 
R ,& 
0 ,& 
0 ,& 
0 ,& 
0 ,& 
0 ,& 
-R ,& 
0 ,& 
0 ,& 
0 ,& 
0 &
) ^{\sf T} ,\\[2mm]
\nu_6 & = \frac{1}{\sqrt R} & 
( &
0 ,& 
0 ,& 
R ,& 
0 ,& 
0 ,& 
0 ,& 
0 ,& 
0 ,& 
0  ,& 
0 ,& 
0 ,& 
0 ,& 
R ,& 
0 ,& 
0 ,& 
0 &
) ^{\sf T} ,\\[2mm]
\nu_7 & = \frac{1}{\sqrt R} & 
( &
0 ,& 
0 ,& 
0 ,& 
0 ,& 
0 ,& 
0 ,& 
0 ,& 
R ,& 
0 ,& 
- R \, e^{2i\varphi} , & 
0 ,& 
0 ,& 
0 ,& 
0 ,& 
0 ,& 
0 , &
) ^{\sf T} \ .
\end{array}
\eeq
The associated metric $\cM_{mn}$ as in \eqref{cM_equation} can be
written compactly as
\beq
\cM_{mn} y^m y^n
= \frac{4z}{R\, \rho} \Big[
e^{i\varphi} \left( R\, y^2 y^4 
- R \, y^6 y^7 
- i \, z \, y^1 y^2\right)
+ e^{-i\varphi} \, y^3 (
\bar \zeta \, y^1
- R \, y^5)
+ \rho \, (y^1)^2
\Big] \ ,
\eeq 
where $y^m$ are  bookkeeping variables.

\bibliography{references}

\providecommand{\href}[2]{#2}\begingroup\raggedright\begin{thebibliography}{10}

\bibitem{Beem:2013sza}
C.~Beem, M.~Lemos, P.~Liendo, W.~Peelaers, L.~Rastelli and B.~C. van Rees,
  \emph{{Infinite Chiral Symmetry in Four Dimensions}},
  \href{http://dx.doi.org/10.1007/s00220-014-2272-x}{\emph{Commun. Math. Phys.}
  {\bf 336} (2015) 1359--1433}, [\href{http://arxiv.org/abs/1312.5344}{{\tt
  1312.5344}}].

\bibitem{Beccaria:2014jra}
M.~Beccaria, C.~Candu and M.~R. Gaberdiel, \emph{{The large N = 4
  superconformal $W_{\infty}$ algebra}},
  \href{http://dx.doi.org/10.1007/JHEP06(2014)117}{\emph{JHEP} {\bf 06} (2014)
  117}, [\href{http://arxiv.org/abs/1404.1694}{{\tt 1404.1694}}].

\bibitem{Gaberdiel:2010pz}
M.~R. Gaberdiel and R.~Gopakumar, \emph{{An $AdS_3$ Dual for Minimal Model
  CFTs}}, \href{http://dx.doi.org/10.1103/PhysRevD.83.066007}{\emph{Phys. Rev.}
  {\bf D83} (2011) 066007}, [\href{http://arxiv.org/abs/1011.2986}{{\tt
  1011.2986}}].

\bibitem{Gaberdiel:2012uj}
M.~R. Gaberdiel and R.~Gopakumar, \emph{{Minimal Model Holography}},
  \href{http://dx.doi.org/10.1088/1751-8113/46/21/214002}{\emph{J. Phys.} {\bf
  A46} (2013) 214002}, [\href{http://arxiv.org/abs/1207.6697}{{\tt
  1207.6697}}].

\bibitem{Gaberdiel:2011zw}
M.~R. Gaberdiel, R.~Gopakumar, T.~Hartman and S.~Raju, \emph{{Partition
  Functions of Holographic Minimal Models}},
  \href{http://dx.doi.org/10.1007/JHEP08(2011)077}{\emph{JHEP} {\bf 08} (2011)
  077}, [\href{http://arxiv.org/abs/1106.1897}{{\tt 1106.1897}}].

\bibitem{Gaberdiel:2011wb}
M.~R. Gaberdiel and T.~Hartman, \emph{{Symmetries of Holographic Minimal
  Models}}, \href{http://dx.doi.org/10.1007/JHEP05(2011)031}{\emph{JHEP} {\bf
  05} (2011) 031}, [\href{http://arxiv.org/abs/1101.2910}{{\tt 1101.2910}}].

\bibitem{Gaberdiel:2011nt}
M.~R. Gaberdiel and C.~Vollenweider, \emph{{Minimal Model Holography for
  SO(2N)}}, \href{http://dx.doi.org/10.1007/JHEP08(2011)104}{\emph{JHEP} {\bf
  08} (2011) 104}, [\href{http://arxiv.org/abs/1106.2634}{{\tt 1106.2634}}].

\bibitem{Candu:2012ne}
C.~Candu, M.~R. Gaberdiel, M.~Kelm and C.~Vollenweider, \emph{{Even spin
  minimal model holography}},
  \href{http://dx.doi.org/10.1007/JHEP01(2013)185}{\emph{JHEP} {\bf 01} (2013)
  185}, [\href{http://arxiv.org/abs/1211.3113}{{\tt 1211.3113}}].

\bibitem{Candu:2012jq}
C.~Candu and M.~R. Gaberdiel, \emph{{Supersymmetric holography on $AdS_3$}},
  \href{http://dx.doi.org/10.1007/JHEP09(2013)071}{\emph{JHEP} {\bf 09} (2013)
  071}, [\href{http://arxiv.org/abs/1203.1939}{{\tt 1203.1939}}].

\bibitem{Creutzig:2012ar}
T.~Creutzig, Y.~Hikida and P.~B. Rønne, \emph{{N=1 supersymmetric higher spin
  holography on $AdS_3$}},
  \href{http://dx.doi.org/10.1007/JHEP02(2013)019}{\emph{JHEP} {\bf 02} (2013)
  019}, [\href{http://arxiv.org/abs/1209.5404}{{\tt 1209.5404}}].

\bibitem{Gaberdiel:2013vva}
M.~R. Gaberdiel and R.~Gopakumar, \emph{{Large N=4 Holography}},
  \href{http://dx.doi.org/10.1007/JHEP09(2013)036}{\emph{JHEP} {\bf 09} (2013)
  036}, [\href{http://arxiv.org/abs/1305.4181}{{\tt 1305.4181}}].

\bibitem{Beccaria:2013wqa}
M.~Beccaria, C.~Candu, M.~R. Gaberdiel and M.~Groher, \emph{{N=1 extension of
  minimal model holography}},
  \href{http://dx.doi.org/10.1007/JHEP07(2013)174}{\emph{JHEP} {\bf 07} (2013)
  174}, [\href{http://arxiv.org/abs/1305.1048}{{\tt 1305.1048}}].

\bibitem{Creutzig:2014ula}
T.~Creutzig, Y.~Hikida and P.~B. Ronne, \emph{{Higher spin AdS$_{3}$ holography
  with extended supersymmetry}},
  \href{http://dx.doi.org/10.1007/JHEP10(2014)163}{\emph{JHEP} {\bf 10} (2014)
  163}, [\href{http://arxiv.org/abs/1406.1521}{{\tt 1406.1521}}].

\bibitem{Candu:2014yva}
C.~Candu, C.~Peng and C.~Vollenweider, \emph{{Extended supersymmetry in
  AdS$_{3}$ higher spin theories}},
  \href{http://dx.doi.org/10.1007/JHEP12(2014)113}{\emph{JHEP} {\bf 12} (2014)
  113}, [\href{http://arxiv.org/abs/1408.5144}{{\tt 1408.5144}}].

\bibitem{Prokushkin:1998bq}
S.~F. Prokushkin and M.~A. Vasiliev, \emph{{Higher spin gauge interactions for
  massive matter fields in 3-D AdS space-time}},
  \href{http://dx.doi.org/10.1016/S0550-3213(98)00839-6}{\emph{Nucl. Phys.}
  {\bf B545} (1999) 385}, [\href{http://arxiv.org/abs/hep-th/9806236}{{\tt
  hep-th/9806236}}].

\bibitem{Henneaux:2010xg}
M.~Henneaux and S.-J. Rey, \emph{{Nonlinear $W_{\infty}$ as Asymptotic Symmetry
  of Three-Dimensional Higher Spin Anti-de Sitter Gravity}},
  \href{http://dx.doi.org/10.1007/JHEP12(2010)007}{\emph{JHEP} {\bf 12} (2010)
  007}, [\href{http://arxiv.org/abs/1008.4579}{{\tt 1008.4579}}].

\bibitem{Campoleoni:2010zq}
A.~Campoleoni, S.~Fredenhagen, S.~Pfenninger and S.~Theisen, \emph{{Asymptotic
  symmetries of three-dimensional gravity coupled to higher-spin fields}},
  \href{http://dx.doi.org/10.1007/JHEP11(2010)007}{\emph{JHEP} {\bf 11} (2010)
  007}, [\href{http://arxiv.org/abs/1008.4744}{{\tt 1008.4744}}].

\bibitem{Campoleoni:2011hg}
A.~Campoleoni, S.~Fredenhagen and S.~Pfenninger, \emph{{Asymptotic W-symmetries
  in three-dimensional higher-spin gauge theories}},
  \href{http://dx.doi.org/10.1007/JHEP09(2011)113}{\emph{JHEP} {\bf 09} (2011)
  113}, [\href{http://arxiv.org/abs/1107.0290}{{\tt 1107.0290}}].

\bibitem{Ito:1992bi}
K.~Ito, J.~O. Madsen and J.~L. Petersen, \emph{{Extended superconformal
  algebras from classical and quantum Hamiltonian reduction}},  in
  \emph{{International Workshop on String Theory, Quantum Gravity and the
  Unification of Fundamental Interactions Rome, Italy, September 21-26, 1992}},
  pp.~302--330, 1992.
\newblock \href{http://arxiv.org/abs/hep-th/9211019}{{\tt hep-th/9211019}}.

\bibitem{deBoer:1998kjm}
J.~de~Boer, \emph{{Six-dimensional supergravity on $S^3 \times AdS_3$ and 2-D
  conformal field theory}},
  \href{http://dx.doi.org/10.1016/S0550-3213(99)00160-1}{\emph{Nucl. Phys.}
  {\bf B548} (1999) 139--166}, [\href{http://arxiv.org/abs/hep-th/9806104}{{\tt
  hep-th/9806104}}].

\bibitem{Minahan:2015jta}
J.~A. Minahan and M.~Zabzine, \emph{{Gauge theories with 16 supersymmetries on
  spheres}}, \href{http://dx.doi.org/10.1007/JHEP03(2015)155}{\emph{JHEP} {\bf
  03} (2015) 155}, [\href{http://arxiv.org/abs/1502.07154}{{\tt 1502.07154}}].

\bibitem{Dabholkar:2010uh}
A.~Dabholkar, J.~Gomes and S.~Murthy, \emph{{Quantum black holes, localization
  and the topological string}},
  \href{http://dx.doi.org/10.1007/JHEP06(2011)019}{\emph{JHEP} {\bf 06} (2011)
  019}, [\href{http://arxiv.org/abs/1012.0265}{{\tt 1012.0265}}].

\bibitem{Dabholkar:2011ec}
A.~Dabholkar, J.~Gomes and S.~Murthy, \emph{{Localization \& Exact
  Holography}}, \href{http://dx.doi.org/10.1007/JHEP04(2013)062}{\emph{JHEP}
  {\bf 04} (2013) 062}, [\href{http://arxiv.org/abs/1111.1161}{{\tt
  1111.1161}}].

\bibitem{Gupta:2012cy}
R.~K. Gupta and S.~Murthy, \emph{{All solutions of the localization equations
  for N=2 quantum black hole entropy}},
  \href{http://dx.doi.org/10.1007/JHEP02(2013)141}{\emph{JHEP} {\bf 02} (2013)
  141}, [\href{http://arxiv.org/abs/1208.6221}{{\tt 1208.6221}}].

\bibitem{Murthy:2013xpa}
S.~Murthy and V.~Reys, \emph{{Quantum black hole entropy and the holomorphic
  prepotential of N=2 supergravity}},
  \href{http://dx.doi.org/10.1007/JHEP10(2013)099}{\emph{JHEP} {\bf 10} (2013)
  099}, [\href{http://arxiv.org/abs/1306.3796}{{\tt 1306.3796}}].

\bibitem{Dabholkar:2014wpa}
A.~Dabholkar, N.~Drukker and J.~Gomes, \emph{{Localization in supergravity and
  quantum $AdS_4/CFT_3$ holography}},
  \href{http://dx.doi.org/10.1007/JHEP10(2014)090}{\emph{JHEP} {\bf 10} (2014)
  90}, [\href{http://arxiv.org/abs/1406.0505}{{\tt 1406.0505}}].

\bibitem{Murthy:2015yfa}
S.~Murthy and V.~Reys, \emph{{Functional determinants, index theorems, and
  exact quantum black hole entropy}},
  \href{http://dx.doi.org/10.1007/JHEP12(2015)028}{\emph{JHEP} {\bf 12} (2015)
  028}, [\href{http://arxiv.org/abs/1504.01400}{{\tt 1504.01400}}].

\bibitem{Gupta:2015gga}
R.~K. Gupta, Y.~Ito and I.~Jeon, \emph{{Supersymmetric Localization for BPS
  Black Hole Entropy: 1-loop Partition Function from Vector Multiplets}},
  \href{http://dx.doi.org/10.1007/JHEP11(2015)197}{\emph{JHEP} {\bf 11} (2015)
  197}, [\href{http://arxiv.org/abs/1504.01700}{{\tt 1504.01700}}].

\bibitem{Murthy:2015zzy}
S.~Murthy and V.~Reys, \emph{{Single-centered black hole microstate
  degeneracies from instantons in supergravity}},
  \href{http://dx.doi.org/10.1007/JHEP04(2016)052}{\emph{JHEP} {\bf 04} (2016)
  052}, [\href{http://arxiv.org/abs/1512.01553}{{\tt 1512.01553}}].

\bibitem{Aharony:2015hix}
O.~Aharony, M.~Berkooz, A.~Karasik and T.~Vaknin, \emph{{Supersymmetric field
  theories on AdS$_{p} \times$ S$^{q}$}},
  \href{http://dx.doi.org/10.1007/JHEP04(2016)066}{\emph{JHEP} {\bf 04} (2016)
  066}, [\href{http://arxiv.org/abs/1512.04698}{{\tt 1512.04698}}].

\bibitem{Assel:2016pgi}
B.~Assel, D.~Martelli, S.~Murthy and D.~Yokoyama, \emph{{Localization of
  supersymmetric field theories on non-compact hyperbolic three-manifolds}},
  \href{http://arxiv.org/abs/1609.08071}{{\tt 1609.08071}}.

\bibitem{Dolan:2002zh}
F.~A. Dolan and H.~Osborn, \emph{{On short and semi-short representations for
  four-dimensional superconformal symmetry}},
  \href{http://dx.doi.org/10.1016/S0003-4916(03)00074-5}{\emph{Annals Phys.}
  {\bf 307} (2003) 41--89}, [\href{http://arxiv.org/abs/hep-th/0209056}{{\tt
  hep-th/0209056}}].

\bibitem{Ademollo:1976pp}
M.~Ademollo et~al., \emph{{Dual String with U(1) Color Symmetry}},
  \href{http://dx.doi.org/10.1016/0550-3213(76)90483-1}{\emph{Nucl. Phys.} {\bf
  B111} (1976) 77--110}.

\bibitem{Witten:1988hf}
E.~Witten, \emph{{Quantum Field Theory and the Jones Polynomial}},
  \href{http://dx.doi.org/10.1007/BF01217730}{\emph{Commun. Math. Phys.} {\bf
  121} (1989) 351--399}.

\bibitem{Moore:1989yh}
G.~W. Moore and N.~Seiberg, \emph{{Taming the Conformal Zoo}},
  \href{http://dx.doi.org/10.1016/0370-2693(89)90897-6}{\emph{Phys. Lett.} {\bf
  B220} (1989) 422--430}.

\bibitem{Elitzur:1989nr}
S.~Elitzur, G.~W. Moore, A.~Schwimmer and N.~Seiberg, \emph{{Remarks on the
  Canonical Quantization of the Chern-Simons-Witten Theory}},
  \href{http://dx.doi.org/10.1016/0550-3213(89)90436-7}{\emph{Nucl. Phys.} {\bf
  B326} (1989) 108--134}.

\bibitem{Pestun:2007rz}
V.~Pestun, \emph{{Localization of gauge theory on a four-sphere and
  supersymmetric Wilson loops}},
  \href{http://dx.doi.org/10.1007/s00220-012-1485-0}{\emph{Commun. Math. Phys.}
  {\bf 313} (2012) 71--129}, [\href{http://arxiv.org/abs/0712.2824}{{\tt
  0712.2824}}].

\bibitem{DiPietro:2015zia}
L.~Di~Pietro, N.~Klinghoffer and I.~Shamir, \emph{{On Supersymmetry, Boundary
  Actions and Brane Charges}},
  \href{http://dx.doi.org/10.1007/JHEP02(2016)163}{\emph{JHEP} {\bf 02} (2016)
  163}, [\href{http://arxiv.org/abs/1502.05976}{{\tt 1502.05976}}].

\bibitem{Gukov:2004id}
S.~Gukov, E.~Martinec, G.~W. Moore and A.~Strominger, \emph{{Chern-Simons gauge
  theory and the $AdS_3/CFT_2$ correspondence}},
  \href{http://arxiv.org/abs/hep-th/0403225}{{\tt hep-th/0403225}}.

\bibitem{Kraus:2006nb}
P.~Kraus and F.~Larsen, \emph{{Partition functions and elliptic genera from
  supergravity}},
  \href{http://dx.doi.org/10.1088/1126-6708/2007/01/002}{\emph{JHEP} {\bf 01}
  (2007) 002}, [\href{http://arxiv.org/abs/hep-th/0607138}{{\tt
  hep-th/0607138}}].

\bibitem{Kraus:2006wn}
P.~Kraus, \emph{{Lectures on black holes and the $AdS_3/ CFT_2$
  correspondence}}, {\emph{Lect. Notes Phys.} {\bf 755} (2008) 193--247},
  [\href{http://arxiv.org/abs/hep-th/0609074}{{\tt hep-th/0609074}}].

\bibitem{Freedman:1998tz}
D.~Z. Freedman, S.~D. Mathur, A.~Matusis and L.~Rastelli, \emph{{Correlation
  functions in the $CFT_d/ AdS_{d+1}$ correspondence}},
  \href{http://dx.doi.org/10.1016/S0550-3213(99)00053-X}{\emph{Nucl. Phys.}
  {\bf B546} (1999) 96--118}, [\href{http://arxiv.org/abs/hep-th/9804058}{{\tt
  hep-th/9804058}}].

\bibitem{Chang:2013fba}
C.-M. Chang and X.~Yin, \emph{{1/16 BPS states in $\mathcal N=$ 4
  super-Yang-Mills theory}},
  \href{http://dx.doi.org/10.1103/PhysRevD.88.106005}{\emph{Phys. Rev.} {\bf
  D88} (2013) 106005}, [\href{http://arxiv.org/abs/1305.6314}{{\tt
  1305.6314}}].

\bibitem{Kim:1985ez}
H.~J. Kim, L.~J. Romans and P.~van Nieuwenhuizen, \emph{{The Mass Spectrum of
  Chiral N=2 D=10 Supergravity on $S^5$}},
  \href{http://dx.doi.org/10.1103/PhysRevD.32.389}{\emph{Phys. Rev.} {\bf D32}
  (1985) 389}.

\bibitem{Breitenlohner:1982jf}
P.~Breitenlohner and D.~Z. Freedman, \emph{{Stability in Gauged Extended
  Supergravity}},
  \href{http://dx.doi.org/10.1016/0003-4916(82)90116-6}{\emph{Annals Phys.}
  {\bf 144} (1982) 249}.

\bibitem{Bowcock:1991zk}
P.~Bowcock and G.~M.~T. Watts, \emph{{On the classification of quantum W
  algebras}}, \href{http://dx.doi.org/10.1016/0550-3213(92)90590-8}{\emph{Nucl.
  Phys.} {\bf B379} (1992) 63--95},
  [\href{http://arxiv.org/abs/hep-th/9111062}{{\tt hep-th/9111062}}].

\bibitem{Beem:2014kka}
C.~Beem, L.~Rastelli and B.~C. van Rees, \emph{{$ \mathcal{W} $ symmetry in six
  dimensions}}, \href{http://dx.doi.org/10.1007/JHEP05(2015)017}{\emph{JHEP}
  {\bf 05} (2015) 017}, [\href{http://arxiv.org/abs/1404.1079}{{\tt
  1404.1079}}].

\bibitem{Chester:2014mea}
S.~M. Chester, J.~Lee, S.~S. Pufu and R.~Yacoby, \emph{{Exact Correlators of
  BPS Operators from the 3d Superconformal Bootstrap}},
  \href{http://dx.doi.org/10.1007/JHEP03(2015)130}{\emph{JHEP} {\bf 03} (2015)
  130}, [\href{http://arxiv.org/abs/1412.0334}{{\tt 1412.0334}}].

\bibitem{Beem:2016cbd}
C.~Beem, W.~Peelaers and L.~Rastelli, \emph{{Deformation quantization and
  superconformal symmetry in three dimensions}},
  \href{http://arxiv.org/abs/1601.05378}{{\tt 1601.05378}}.

\bibitem{Costello:2016mgj}
K.~Costello and S.~Li, \emph{{Twisted supergravity and its quantization}},
  \href{http://arxiv.org/abs/1606.00365}{{\tt 1606.00365}}.

\end{thebibliography}\endgroup
\bibliographystyle{JHEP}

\end{document}